\documentclass[twocolumn]{aastex62}

\usepackage{hyperref}

\newcommand\teff{$T_{\mathrm{eff}}$}
\newcommand\logg{$\log g$}

\newcommand{\kep}{\mbox{\textit{Kepler}}}
\newcommand{\gaia}{\mbox{\textit{Gaia}}}

\newcommand{\bt}{$B_{\mathrm{T}}$}
\newcommand{\bteqn}{B_{\mathrm{T}}}
\newcommand{\vt}{$V_{\mathrm{T}}$}
\newcommand{\vteqn}{V_{\mathrm{T}}}

\newcommand{\mdot}{$\mathrm{M_\odot}$}
\newcommand{\rdot}{$\mathrm{R_\odot}$}
\newcommand{\ldot}{$\mathrm{L_\odot}$}

\newcommand{\nstarsinput}{\mbox{186,548}}
\newcommand{\nstars}{\mbox{186,301}}

\usepackage{amsmath}

\graphicspath{{./}{figures/}}

\shorttitle{Gaia-Kepler Stellar Properties Catalog}
\shortauthors{Berger et al.}

\begin{document}

\title{The $Gaia$-$Kepler$ Stellar Properties Catalog. I. Homogeneous Fundamental Properties for 186,301 \kep\ Stars}

\correspondingauthor{Travis Berger}
\email{taberger@hawaii.edu}

\author[0000-0002-2580-3614]{Travis A. Berger}
\affiliation{Institute for Astronomy, University of Hawai`i, 2680 Woodlawn Drive, Honolulu, HI 96822, USA}

\author[0000-0001-8832-4488]{Daniel Huber}
\affiliation{Institute for Astronomy, University of Hawai`i, 2680 Woodlawn Drive, Honolulu, HI 96822, USA}

\author[0000-0002-4284-8638]{Jennifer L. van Saders}
\affiliation{Institute for Astronomy, University of Hawai`i, 2680 Woodlawn Drive, Honolulu, HI 96822, USA}

\author[0000-0002-5258-6846]{Eric Gaidos}
\affiliation{Department of Earth Sciences, University of Hawai`i at M\={a}noa, Honolulu, HI 96822, USA}

\author[0000-0002-4818-7885]{Jamie Tayar}\altaffiliation{Hubble Fellow}
\affiliation{Institute for Astronomy, University of Hawai`i, 2680 Woodlawn Drive, Honolulu, HI 96822, USA}

\author[0000-0001-9811-568X]{Adam L. Kraus}
\affiliation{Department of Astronomy, The University of Texas at Austin, Austin, TX 78712, USA}

\begin{abstract}
\noindent
An accurate and precise \kep\ Stellar Properties Catalog is essential for the interpretation of the \kep\ exoplanet survey results. Previous \kep\ Stellar Properties Catalogs have focused on reporting the best-available parameters for each star, but this has required combining data from a variety of heterogeneous sources. We present the \gaia-\kep\ Stellar Properties Catalog, a set of stellar properties of \nstars\ \kep\ stars, homogeneously derived from isochrones and broadband photometry, \gaia\ Data Release 2 parallaxes, and spectroscopic metallicities, where available. Our photometric effective temperatures, derived from $g-K_s$ colors, are calibrated on stars with interferometric angular diameters. Median catalog uncertainties are 112\,K for \teff, 0.05\,dex for \logg, 4\% for $R_\star$, 7\% for $M_\star$, 13\% for $\rho_\star$, 10\% for $L_\star$, and 56\% for stellar age. These precise constraints on stellar properties for this sample of stars will allow unprecedented investigations into trends in stellar and exoplanet properties as a function of stellar mass and age. In addition, our homogeneous parameter determinations will permit more accurate calculations of planet occurrence and trends with stellar properties.
\end{abstract}

\section{Introduction} \label{sec:intro}

The \kep\ Mission, officially retired in 2018, has left an unprecedented legacy dataset for stellar astrophysics and exoplanet science. Due to the long baseline, high precision observations and subsequent follow-up efforts, the \kep\ target stars have become one of the best-characterized samples of stars \citep{huber14,Mathur2017,Berger2018c}.

The original \kep\ Input Catalog \citep[KIC,][]{brown11} compiled data for the purpose of target selection. It included optical photometry ($griz$), \teff, \logg, and metallicities. From the $\sim$\,13 million stars within the KIC, $\sim$\,200,000 stars were chosen for monitoring based on the KIC stellar properties. The exact selection function is complex, but solar-type stars were prioritized according to more precise determinations of the \kep\ sample's stellar properties \citep{batalha10,Berger2018c}. Overall, these $\sim$\,200,000 target stars either had imprecise stellar parameters -- 0.3--0.4\,dex uncertainties in \logg\ \citep{brown11,huber16} and $\approx$\,200\,K uncertainties in \teff\ -- or lacked parameters altogether, such as masses, ages, radii, densities, and distances.

The first \kep\ Stellar Properties Catalog \citep[KSPC,][]{huber14} was published to consolidate all of the follow-up work done for \kep\ stars and improve the estimated planetary occurrence rates \citep[e.g.,][]{Howard2012,petigura13,burke15,Fulton2017}. This catalog included follow-up spectroscopy, spectroscopic surveys, and asteroseismic analysis. In addition to the 12,000 \kep\ stars with asteroseismic constraints prior to 2014, \cite{huber14} analyzed another $\sim$\,3000 oscillating stars, providing a total of $\approx$\,15,500 stars with asteroseismic radii and masses. However, radii and masses for most stars remained imprecise due to the vast majority of stars having only photometric constraints.

In the years following the first KSPC, the number of \kep\ stars with spectroscopic constraints increased considerably due to two large scale spectroscopic surveys: (1) the Apache Point Observatory for Galactic Evolution Experiment \citep[APOGEE,][]{Majewski2017} and (2) the Large Sky Area Multi-Object Fiber Spectroscopic Telescope survey \citep[LAMOST,][]{Zhao2012,luo15}. \cite{Mathur2017} implemented this spectroscopy in addition to \logg\ constraints from the stellar granulation-driven flicker method \citep{Bastien2016} to produce the Data Release 25 (DR25) KSPC. In combination with improved methodology, these additional data led to typical uncertainties of $\approx$\,27\% in radius, $\approx$\,17\% in mass, and $\approx$\,51\% in density. The large median catalog uncertainties on radius and density remained due to a lack of additional data (e.g. spectroscopy and parallaxes) for the majority of stars.

Fortunately, \gaia\ DR2 \citep{Arenou2018,Brown2018,Lindegren2018} recently provided parallaxes to 1.3 billion stars, including $<$\,20\% parallaxes for $\sim$\,180,000 \kep\ stars \citep{Berger2018c}. Combining \gaia\ DR2 parallaxes with \teff\ and $K_s$-band magnitudes from the DR25 KSPC \citep{Mathur2017}, \cite{Berger2018c} recomputed stellar radii and luminosities for 177,911 \kep\ stars, updating our census of the \kep\ targets with median radius precisions of 8\% and allowing us to determine the fraction of main sequence (67\%), subgiant (21\%) and giant (12\%) stars in the \kep\ target list. However, this work did not provide masses, \logg, or densities because isochrones were not used. Isochrones are required to derive physical parameters such as mass, \logg, and density from bulk observables such as parallaxes and photometry.

In this paper, we utilize \gaia\ DR2 parallaxes, homogeneous stellar $g$ and $K_s$ photometry, and spectroscopic metallicities, where available, to improve on previous analyses and present the most accurate, homogeneous, and precise analysis of stars in the \kep\ field. We re-derive stellar \teff, \logg, radii, masses, densities, luminosities, and ages for \nstars\ \kep\ targets, and investigate the stellar properties of a number of noteworthy \kep\ exoplanet-hosting stars.

\section{Methodology} \label{sec:methods}

\subsection{Sample Selection}\label{sec:flagscuts}

To identify our sample, we use the same \gaia-\kep\ cross-match detailed in \cite{Berger2018c}, which included 195,710 stars. We removed stars lacking ``AAA'' 2MASS photometry \citep{skrutskie06} and any stars lacking measured parallaxes in \gaia\ DR2. Requiring ``AAA'' 2MASS photometry means we removed the brightest stars due to saturation and the faintest stars due to photon noise. In addition, after crossmatching our sample with the binaries of \cite{Kraus2016}, we found that 248 out of the 263 matched primaries had ``AAA'' 2MASS photometry, while seven of those without ``AAA'' photometry had low contrast ($\Delta m$\,$\lesssim$\,2\,mag), moderately resolved (2''\,$<$\,$\theta_{\mathrm{sep}}$\,$<$\,4'') companions. Stars lacking parallaxes are typically moderately close ($\sim$\,200--400\,mas) equal brightness binaries, whereas other binaries at least have parallaxes, even if the errors are larger (Kraus et al. in prep). These cuts reduced the sample to 190,213 stars and then to 186,672 stars, successively. Requiring $g$-band photometry from either the KIC or the \kep-INT Survey \citep[KIS,][]{Greiss2012} reduced the catalog to 186,548 stars.

\subsection{Input Photometry}\label{sec:inputphotometry}

\begin{figure}
\resizebox{\hsize}{!}{\includegraphics{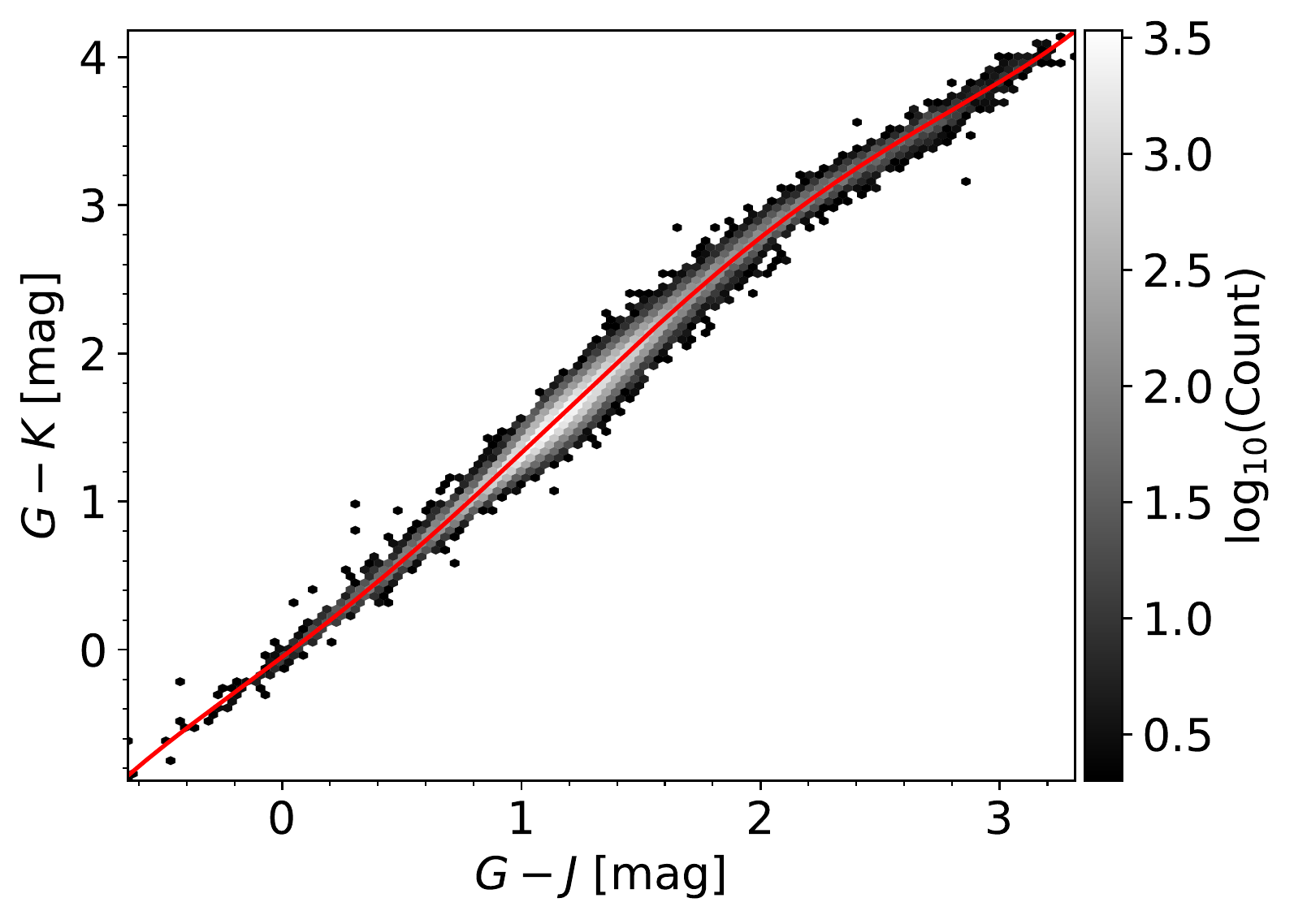}}
\resizebox{\hsize}{!}{\includegraphics{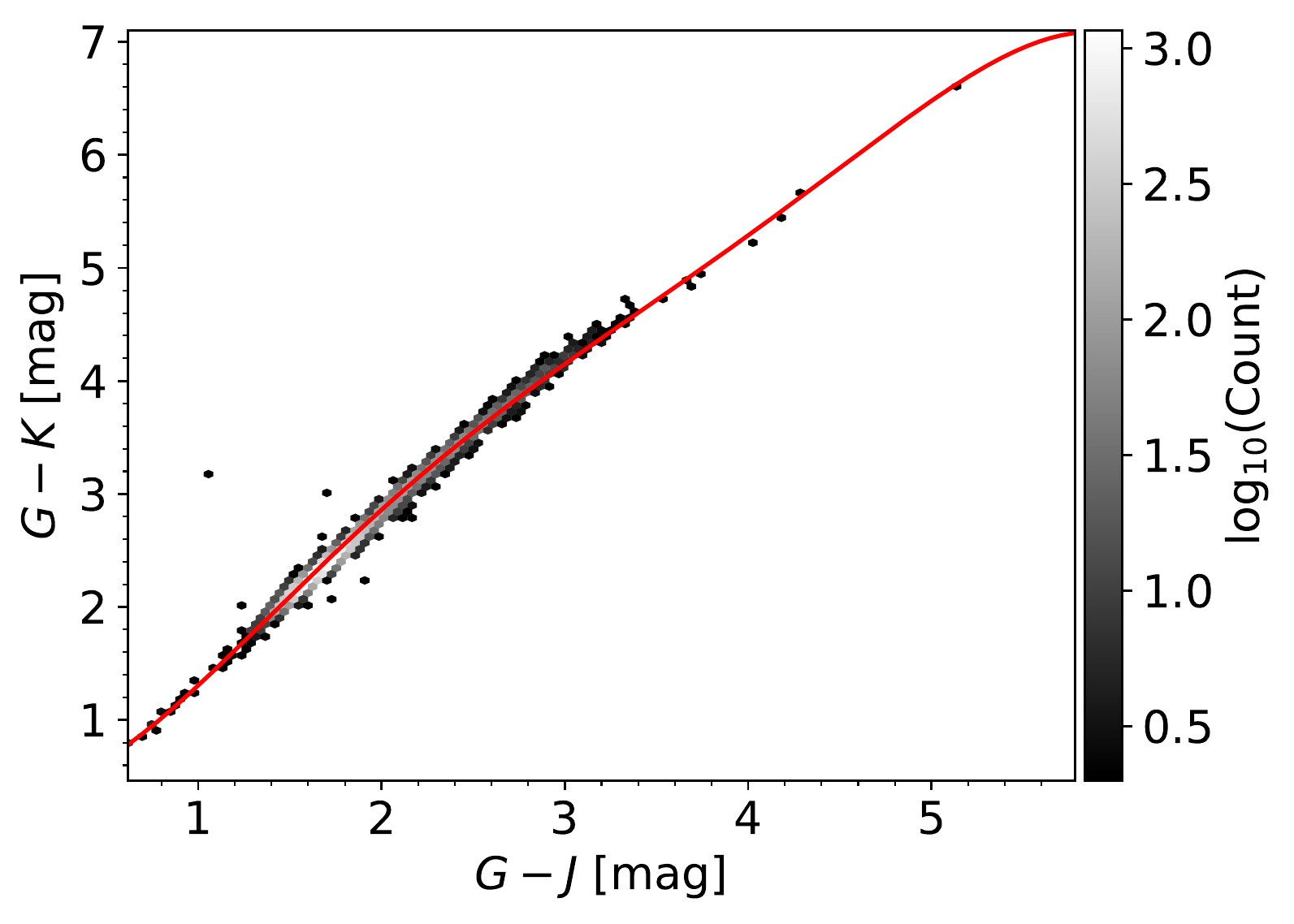}}
\caption{$G-K_s$ vs $G-J$ of all stars without \gaia\ companions within 4'' for dwarfs (top panel) and giants (bottom panel). Color-coding represents logarithmic number density. The red line displays the best-fit fifth-order polynomial to the locus of points.}
\label{fig:Kpoly}
\end{figure}

To create a homogeneous catalog for the entire \kep\ target sample, we mainly used Sloan $g$ and 2MASS $K_s$ photometry. We chose these two passbands to maximize both our \teff\ sensitivity and the number of stars included in our final catalog. We avoid using additional SDSS or other ground-based bandpasses which would further reduce our sample. However, we do use $J$ magnitudes to derive estimated $K_s$ magnitudes for binary secondaries, although they are not used directly as an input to our isochrone fitting process. While \gaia\ $G$, $B_p$, and $R_p$ are available for the vast majority of \kep\ stars, there remains ongoing work to sufficiently characterize their transmission profiles for synthetic photometry.

We took our $K_s$ photometry solely from 2MASS, which has an effective angular resolution of 4'', where any fainter sources between 1.5'' and 4'' were omitted \citep{skrutskie06}. We then used \gaia\ DR2 photometry, which has a resolution of $\lesssim$\,1'', to identify contaminating sources within 4''. To do this, we first cross-matched \gaia\ DR2 sources within 4'' of our catalog of \kep\ stars. Next, we cross-matched \gaia-detected secondary sources with the United Kingdom Infrared Telescope (UKIRT) $J$-band observations using the WFCAM Science Archive and the WSERV4v20101019 database. We chose a matching radius of 0.5'' from the \gaia\ secondaries based on the minimum in the distribution of angular separation of matches. We do not use these $J$ magnitudes in our isochrone fitting procedure, as $K_s$ magnitudes are less affected by extinction than UKIRT $J$ and maximize our \teff\ sensitivity.

Without taking any additional steps, we might wrongly confuse some UKIRT $J$-band magnitudes as belonging to the secondaries when they belong to the primaries. To find these false matches, we plotted the distribution of $J_{\mathrm{2MASS}} - J_{\mathrm{UKIRT}}$ and found two peaks:  (1) one narrow peak occurring at 0.0\,$\pm$\,0.2\,mag and (2) a broader peak occurring at 3\,$\pm$1\,mag. The first peak corresponds to false secondary-as-primary identifications while the second peak represents true secondaries. Therefore, we excluded all secondaries that have both $|J_{\mathrm{2MASS}} - J_{\mathrm{UKIRT}}|$\,$<$\,0.2\,mag and angular separations $<$\,1.5''.

Figure \ref{fig:Kpoly} displays our computed fifth-order polynomial fits to the $G - K_s$ versus $G - J_{\mathrm{2MASS}}$ curves of the non-binary \kep\ dwarfs (top) and giants (bottom); stars are designated by their evolutionary state \citep[Table 1,][]{Berger2018c}. We removed all stars with binary flags $>$\,0 in Table 1 of \cite{Berger2018c} as well as those with \gaia-detected companions to avoid contaminated secondary $K_s$ magnitudes. We computed secondary $K_s$ magnitudes using the difference of the \gaia\ $G$ magnitude and a $G-K_s$ color. The $G-K_s$ color was computed from the best-fit polynomial evaluated at $G - J_{\mathrm{UKIRT}}$, where $J_{\mathrm{UKIRT}}$ is the ``jAperMag3'' UKIRT $J$ photometry. For secondaries without a UKIRT $J$ magnitude, we did not compute a secondary $K_s$ magnitude.

To compute amended $K_s$ magnitudes for those primary stars with \gaia\ DR2 companions within 4'', we used the following expression:
\begin{equation}
    K_{\mathrm{prim}} = K_{\mathrm{sec}} - 2.5\log(10^{-0.4*(K_{\mathrm{2MASS}} - K_{\mathrm{sec}})} - f(\theta))
\end{equation}
where $f(\theta)$ =
$\begin{cases}
    1.0 & \theta\leq 1.25'' \\
    -\frac{4}{9}\theta + \frac{14}{9} & 1.25 < \theta < 3.5'', \\
    0.0 & \theta \geq 3.5'' \\
\end{cases}$
\vspace{0.3cm}

\noindent and $\theta$ is the angular separation between the primary and the secondary in arcseconds, $K_{\mathrm{prim}}$ is the corrected primary $K_s$ magnitude, $K_{\mathrm{sec}}$ is the secondary $K_s$ magnitude computed as described above, and $K_{\mathrm{2MASS}}$ is the original magnitude provided by 2MASS. The expression for $f$ above represents the fraction of flux of fainter sources contained within the 2MASS aperture. This expression was derived from a comparison of the \cite{Kraus2007} point spread function (PSF) fit magnitudes to the 2MASS catalog \citep{skrutskie06} magnitudes. The comparison reveals that $<$\,1.25'' and $>$\,3.5'' are the places where 100\% and 0\% of the flux appear to be captured, and we use a linear relationship both for simplicity and because it is consistent with the binary fitting results of \cite{Kraus2007}.

We provide the status of these corrections in Table \ref{tab:input}, flagging \kep\ stars with one \gaia\ resolved companion but no correction as ``BinDetNoCorr'' and those with one resolved companion with a correction as ``BinaryCorr''. Stars with multiple resolved companions are flagged as ``TerDetNoCorr'', ``TerDetBinCorr'', and ``TertiaryCorr'' depending on whether we computed $K_s$-magnitude corrections for zero, one, or multiple companions, respectively. We also include the number of companions as an additional column in the table. Finally, we adopted the photometric errors reported by 2MASS.

Unlike the $K_s$ photometry, both the KIC and KIS $g$ photometry require calibration. To convert the KIC $g$ photometry to the SDSS $g$ of the MESA Isochrones and Stellar Tracks \citep[MIST v1.2,][]{choi16,dotter16,paxton11,paxton13,paxton15} grid, we used Equation (1) in \cite{Pinsonneault2012}. We solved the SDSS-Vega system equations provided in Section 4 of \cite{Gonzalez2011} for $g_{\mathrm{SDSS}}$ to convert the Vega system KIS photometry back to the SDSS AB system.

Next, we used KIS photometry for all sources where it was both available and the corresponding photometry flag indicated neither saturation nor bad pixels. We used the calibrated KIC photometry for all other stars. Our final input catalog utilizes KIS $g$ photometry for 148,410 stars, and KIC photometry for the remaining 38,138 stars. We did not amend the $g$-band magnitudes for contamination from secondaries like we did for $K_s$ magnitudes, due to the $\approx$1.5'' effective resolutions of the KIC and KIS catalogs.

Neither \cite{brown11} nor \cite{Greiss2012} report uncertainties for individual sources; hence, we computed the errors of our $g$ magnitudes by utilizing the photometric scatter relations provided in \cite{brown11} and \cite{Greiss2012}:
\begin{equation}
    \sigma_{g_{\mathrm{KIS}}} = \sqrt{0.02^2 + (1.1 * e^{0.6456*g_{\mathrm{KIS}}-16.1181})^2}
\end{equation}
\begin{equation}
    \sigma_{g_{\mathrm{KIC}}} = \sqrt{0.02^2 + (0.01 * (g_{\mathrm{KIC}} - 12))^2}.
\end{equation}
We report our $g$ and $K_s$ photometry and their errors in Table \ref{tab:input}.

\subsection{Isochrone Fitting}\label{sec:modelgrid}

Isochrone fitting allows the straightforward determination of stellar parameters, such as \teff, masses, and ages, from a set of input observables. We used the most recent MIST models \citep[v1.2 with rotation,][]{choi16,dotter16,paxton11,paxton13,paxton15}, which we have interpolated from the grids provided on the MIST website. Our final grid contains $\sim$\,7 million models with 117 ages from 0.1\,Gyr to 3.68\,Gyr in 52 logarithmic steps and 3.68\,Gyr to $\approx$\,20\,Gyr in linear steps of 0.25\,Gyr to sufficiently sample both young and old stellar models and avoid preferential ``snapping" to sparse bands of the model grid. We chose 20\,Gyr as our maximum age because it is the largest age in the MIST grid and it minimizes posterior truncation. Posterior truncation produces an under/overestimation of derived stellar parameters and an underestimation of their errors; this deleterious effect is inevitable with any finite grid, but we minimized its magnitude by including models up to the grid's maximum age. We also do not use any pre-main sequence (PMS) models. The grid has initial metallicities ranging between --2.0 and 0.5\,dex with 0.05\,dex steps. The grid also accounts for element diffusion, which affects the surface abundances.

\begin{figure}
\resizebox{\hsize}{!}{\includegraphics{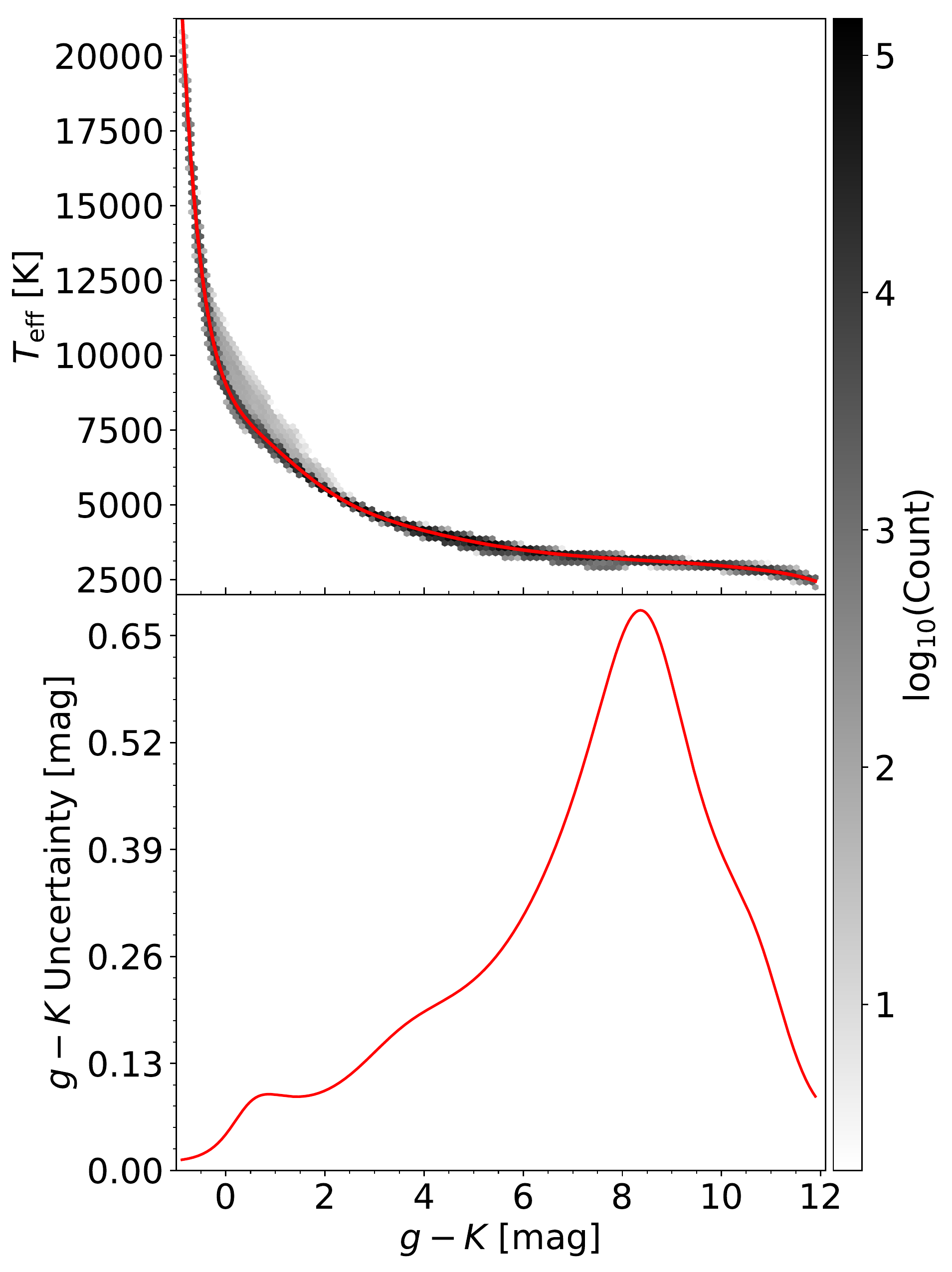}}
\caption{Corrections to the initial $g - K_s$ magnitude errors utilizing our interpolated MIST model grid. $Top$: \teff\ versus $g - K_s$ color for our model grid, with the logarithmic density of points illustrated by the two-dimensional greyscale histogram with the corresponding colorbar. We plot the best-fit 12th order polynomial in red. $Bottom$: The red curve represents the required $g - K_s$ minimum uncertainty to reach a 2\% \teff\ error for all stars, dependent on their $g - K_s$ color.} 
\label{fig:gkerr}
\end{figure}

Isochrone grids frequently struggle to reproduce empirical constraints from M-dwarfs due in part to the presence of starspots and strong magnetic fields \citep{boyajian12,Feiden2012,Mann2019}. Therefore, we implemented the empirical \cite{mann15} and \cite{Mann2019} $r - J$--\teff, $M_{K_s}$--radius and $M_{K_s}$--mass and metallicity relations to compute \teff, stellar radius, and stellar mass, respectively. These relations are mainly calibrated on absolutely flux-calibrated spectra (\teff, radius) and the binary orbital parameters (mass) of nearby M-dwarfs. We do not extrapolate the \cite{mann15} and \cite{Mann2019} relations and only change models that are within the empirical fits:  $1.9 < r - J < 5.5$\,mag, $M_{K_s}$\,$>$\,4.0\,mag, and [Fe/H]\,$>$\,--0.6 dex. Given these photometric and metallicity constraints, we applied the Mann empirical relations to stars with masses below $\approx$\,0.75\,\mdot; hence, we revised $\approx$\,178,000 models. Redder M-dwarf models ($r - J$\,$>$\,5.5 or \teff\,$\lesssim$\,2800\,K) that would require extrapolation are dropped from our grid altogether.

Utilizing solely the individual $g$ and $K_s$ photometric errors added in quadrature yielded \teff\ fractional errors of $\lesssim$\,1\%. These errors are too small given that there are systematic errors in interferometric angular diameters which, in turn, set the fundamental limit on \teff\ errors:  $\approx$\,2\%. Therefore, we computed the best-fitting 12th-order polynomial to the relationship of \teff\ to the $g-K_s$ color of all models within our model grid using \texttt{numpy}'s \texttt{polyfit} routine (Figure \ref{fig:gkerr}). We chose a 12th-order polynomial because all lower-order polynomials do not accurately trace the center of the \teff-color curve, while higher-order polynomials minimally improve the resulting correlation coefficient. For our eventual isochrone fitting, we adopted the maximum $g-K_s$ error:  either (1) the $g$ and $K_s$ errors added in quadrature or (2) the 2\% \teff\ curve-computed value of the input $g - K_s$ color in the bottom panel of Figure \ref{fig:gkerr}. We also found that we underestimate our stellar mass errors for M-dwarfs because of the tight \cite{Mann2019} $M_{K_s}$--mass relation. Hence, we inflate our M-dwarf ($g - K_s$\,$>$\,4, $M_{K_s}$\,$>$\,4) $M_{K_s}$ uncertainties by adding an error term in quadrature corresponding to a 2.1\% mass error, identical to the scatter in the empirical relation \citep{Mann2019}.

We employed \texttt{isoclassify} \citep{Huber2017} and the \cite{Green2019} reddening map to derive stellar parameters from our input observables:  (1) SDSS $g$ and 2MASS $K_s$ photometry, (2) \gaia\ DR2 parallaxes \citep{gaia1,Brown2018,Lindegren2018,Arenou2018}, (3) red giant evolutionary state flags \citep[RGB versus the Red Clump,][]{vrard16,Hon2018}, and (4) spectroscopic metallicities from the \kep\ Stellar Properties Catalog \citep{Mathur2017}, California-\kep\ Survey \citep[CKS,][]{Petigura2017}, APOGEE DR14 \citep{Majewski2017,APOGEEDR14}, and LAMOST DR5 \citep{Zhao2012,LAMOSTDR5}, where available. We computed the 16, 50, and 84th percentile values for \teff, \logg, [Fe/H], radius, mass, density, age, distance, and $V$-band extinction from the marginalized posteriors.

We used conservative 0.15\,dex metallicity errors instead of the quoted errors for all stars with spectroscopic metallicity constraints because the pipeline-to-pipeline uncertainty in metallicities is $\gtrsim$\,0.1\,dex \citep{Furlan2018}. For stars that do not have spectroscopic metallicity constraints ($\sim$\,120,000), we used a prior centered on solar metallicity with a standard deviation of $\sim$\,0.20\,dex, which is appropriate for the \kep\ field \citep{dong14}. As demonstrated in \cite{Howes2019}, a lack of a metallicity constraint typically results in an age-metallicity degeneracy depending on the stellar properties/evolutionary state.

\subsection{Accounting for Binaries}\label{sec:binaries}

As we described in \S\ref{sec:inputphotometry}, we addressed \gaia-detected stellar companions within 4'' which contaminate the 2MASS $K_s$ photometry. In addition, \gaia\ DR2 astrometric flags appear to be another useful tool for identifying binaries that are not resolved by the satellite. For instance, \cite{Evans2018}, \cite{Rizzuto2018}, and \cite{Ziegler2018} have already demonstrated that these astrometric flags are useful for identifying smaller separation binaries that are not spatially resolved.

To identify these smaller separation binaries, we computed \gaia\ DR2's re-normalized unit-weight error (RUWE) by interpolating the tabular data detailed in \cite{RUWE}. RUWE is the magnitude- and color-independent re-normalization of the astrometric $\chi^2$ of \gaia\ DR2 (unit-weight error or UWE). The RUWE values are reported for all \kep\ stars in the ``RUWE'' column of Table \ref{tab:input}. Any stars with RUWE\,$\gtrsim$\,1.2 are likely to be binaries (A. Kraus et al., in prep). Of the 186,548 stars remaining, 170,845 had RUWE\,$<$\,1.4 and 164,736 had RUWE\,$<$\,1.2; we provide RUWE values for every star, but analyze both those with high RUWE and low RUWE similarly. Any derived stellar parameters for these high RUWE stars should be treated with extra caution, along with those that have ``NoCorr'' in their $K_s$ magnitude flags also in Table \ref{tab:input}. We also did not amend the input magnitudes of \kep\ planet host stars with adaptive optics (AO)-detected stellar companions \citep{Furlan2017} in order to preserve the homogeneity of our catalog.

\section{Validating the Output Stellar Parameters}\label{sec:valid}

\subsection{Accuracy of Derived Effective Temperatures}\label{sec:teff}

\begin{figure}
\resizebox{\hsize}{!}{\includegraphics{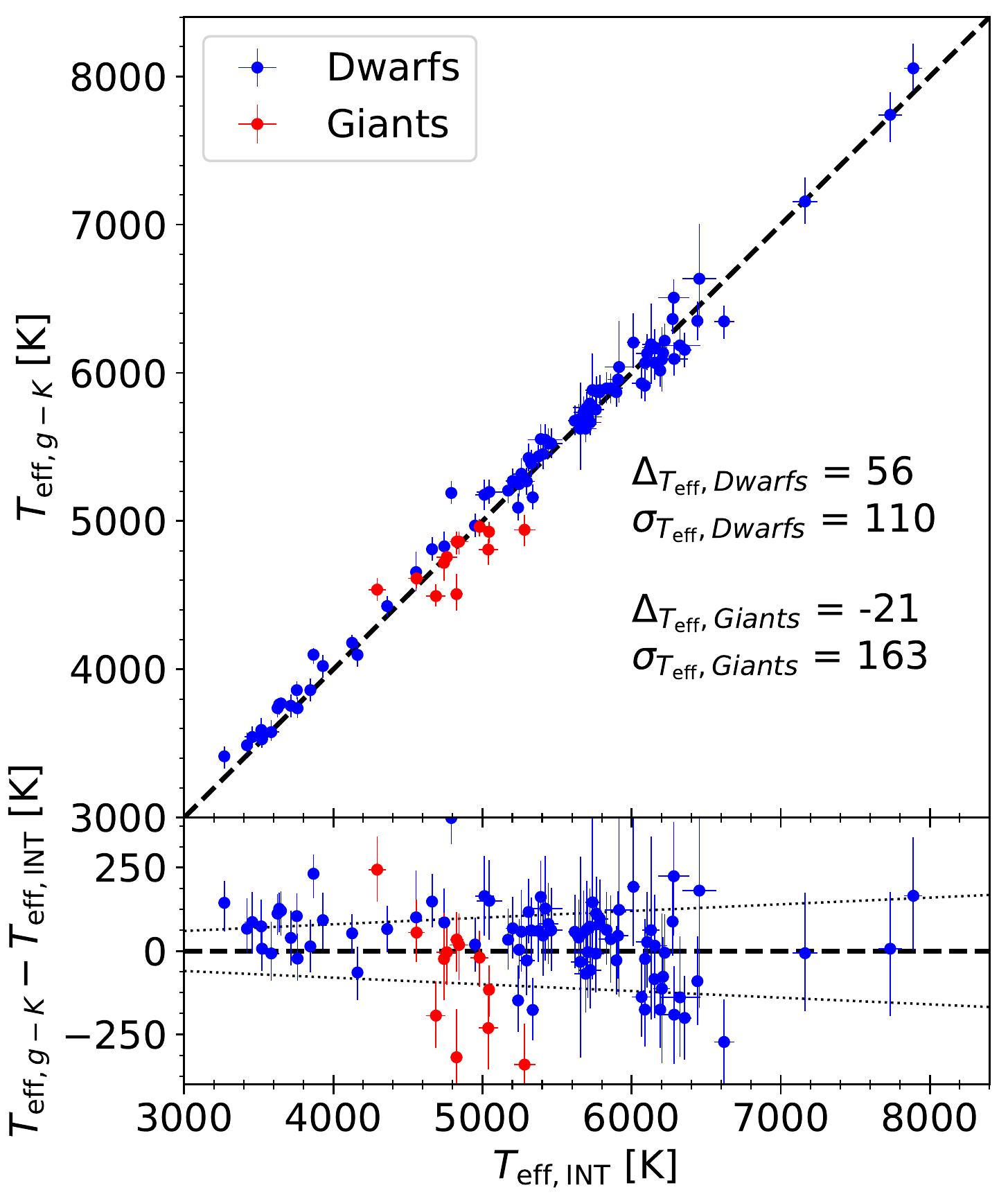}}
\caption{$Top$: Grid-modeled \teff\ versus interferometrically-derived \teff\ \citep{boyajian13,huang15b}. We plot the dwarfs as blue points and giants as red points, as well as their respective uncertainties. The black dashed line is the 1:1 line. The text in the plot indicates the median shifts ($\Delta$) and the median absolute deviations ($\sigma$) for both the dwarfs and giants, as labeled. $Bottom$: Residuals as a function of interferometric \teff. The black dotted lines represent 2\% fractional uncertainties above and below equality.} 
\label{fig:teffint}
\end{figure}

To ensure our grid-computed stellar effective temperatures are accurate, we compared them to interferometric \teff\ measurements for a sample of 108 stars from \cite{boyajian13} and \cite{huang15b} with Tycho $B$ and $V$ photometry as well as 2MASS $K_s$ photometry. Although these stars do have $g$-band photometry from the American Association of Variable Star Observers Photometric All-Sky Survey \citep[APASS,][]{APASS}, it is saturated for these stars. Therefore, we had to convert the Tycho $B$ and $V$ photometry into SDSS $g$ using the following procedure. First, we converted Tycho $B$ (\bt) and $V$ (\vt) photometry into Johnson $B$ and $V$ photometry using the Table 2 from \cite{bessell00}. Then we converted our $B$ and $V$ magnitudes into $g$ magnitudes with the transformation given in the bottom portion of Table 1 of \cite{Jester2005}. We found color-dependent systematics in our comparison of $g$ and \vt\ magnitudes, which we eliminated utilizing the Tycho $\bteqn - \vteqn$ colors. Hence, we computed our interferometric sample $g$ magnitudes as follows:
\begin{equation}
\begin{split}
g\left(B,V,\bteqn,\vteqn\right) = V - 0.03 + 0.60\left(B - V\right)\\
 - 0.11\left(\bteqn - \vteqn\right) + 0.09 \left(\bteqn - \vteqn\right)^2.
\end{split}
\end{equation}

We present the comparison to the interferometric stars in Figure \ref{fig:teffint}. Based on the reported median shifts and median absolute deviations in \teff\ for the dwarfs and giants, our derived \kep\ stellar \teff\ appear to be accurate within our 2\% \teff\ errors for the \teff\ range covered here. We found that the residuals of stars with $K_s$-band errors $>$\,0.25 mag were particularly discrepant in their residuals, and ignore them here. For solar \teff\ and late-G and early-K-dwarfs, we estimate hotter \teff\ than interferometric determinations by 50--60\,K, while we underestimate the \teff\ of F-dwarfs by $\approx$\,100\,K, also within our reported errors. We caution that the derived effective temperatures for M-dwarfs are systematically offset by $\approx$\,75\,K, and the giants appear to demonstrate a strong trend with interferometric \teff. The trend in the giants is likely due to insufficient color transformations and/or saturated photometry, as all interferometric stars are close to 2MASS saturation due to their proximity. Upcoming work on M-dwarfs (Gaidos et al., in prep) and previously published work on the APOGEE Kepler Giants with asteroseismic data \citep[APOKASC,][]{Serenelli2017,Pinsonneault2018} are more specialized and hence better alternatives to the data we derive here for these specific samples of stars.

\subsection{Binary Effects on Stellar Properties}\label{sec:binagefx}

\begin{figure*}
\resizebox{\hsize}{!}{\includegraphics{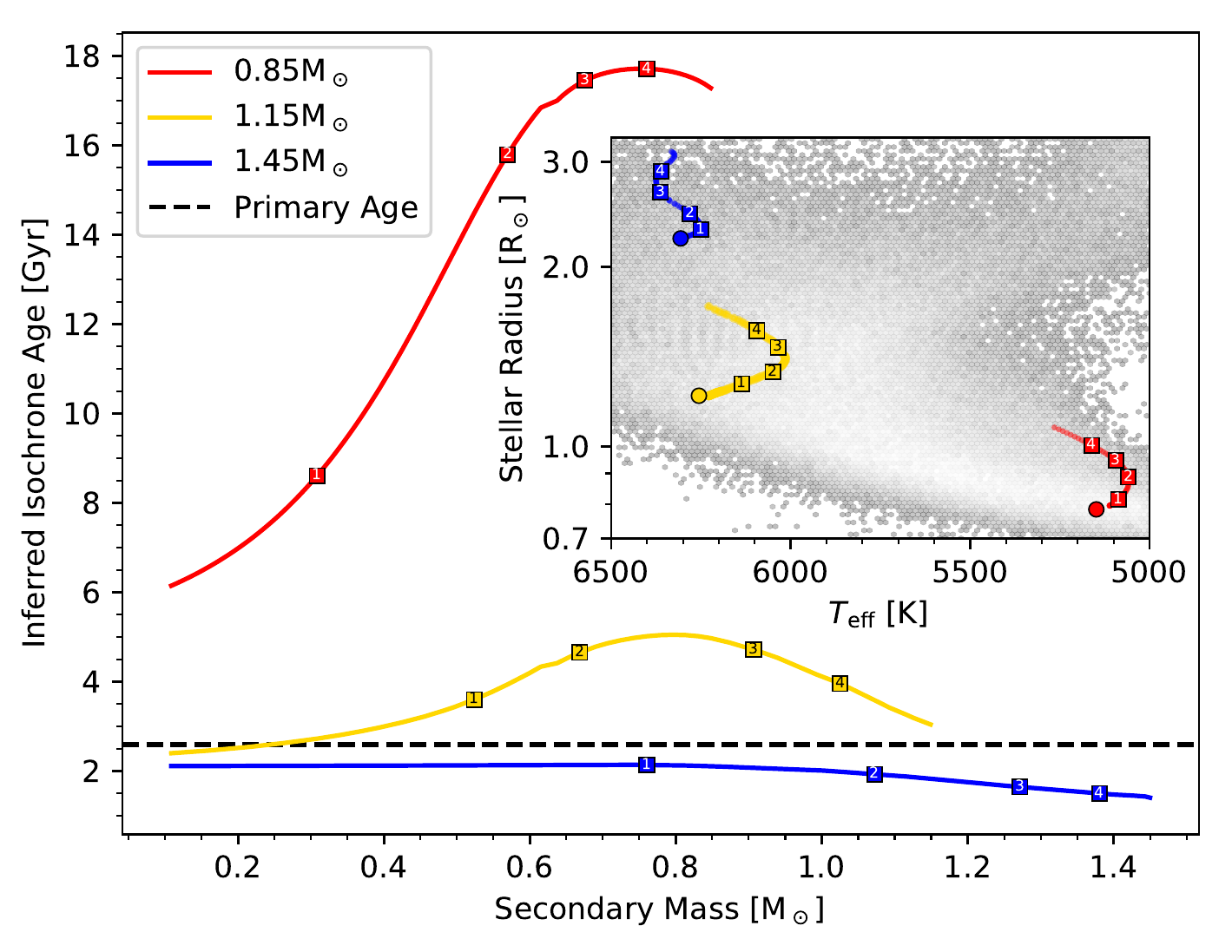}}
\caption{\texttt{isoclassify}-derived age versus the secondary's mass for simulated binary systems with primary masses 0.85\,\mdot\ (red), 1.15\,\mdot\ (yellow), and 1.45\,\mdot\ (blue). Both the primary and the secondaries are 2.59\,Gyr old (black dashed line) and have initial surface metallicities of 0.0\,dex. The inset shows where the sole primary and primary + secondary combinations occur on the H-R diagram. The large, color-matched circles represent the primary's properties and the colored curves are the primary + secondary composite parameters. The color and number-coordinated squares represent the same solutions on both diagrams. To guide the eye, we plot a 2-dimensional histogram for the entire \kep\ catalog underneath, where darker greys represent areas with lower logarithmic number density.}
\label{fig:binagefx}
\end{figure*}

Regardless of how much we account for binaries in our above analysis, there will inevitably be systems that still have unresolved, unidentified stellar companions. These companions will be at angular separations less than both the \gaia\ photometric resolution and RUWE effective resolution, and they will affect our estimates of the stellar properties. To quantify these effects, we took a set of models from our MIST grid of isochrones (\S\ref{sec:modelgrid}) at a particular age and initial metallicity and combined their photometry with that from all stellar models less than or equal to the mass of the primary in that same isochrone. We then ran the photometry for each of these modified models through \texttt{isoclassify}. We used typical \kep\ field \gaia\ DR2 parallaxes and errors and primary surface metallicities, assuming zero reddening for simplicity. We chose primary stars of 0.85, 1.15, and 1.45\,\mdot, at the mode of ages (2.59\,Gyr) and metallicities (0.0\,dex) determined for the \kep\ target sample.

In Figure \ref{fig:binagefx}, we plot the effect of binarity on the age, radius, and \teff\ estimates for our representative set of age, metallicity, and stellar masses for \kep\ stars. In red, yellow, and blue are the binary fitting results for the 0.85\,\mdot, 1.15\,\mdot, and 1.45\,\mdot\ primaries, respectively.

We see that the age of the 0.85\,\mdot\ star is severely affected by the addition of a secondary companion:  even with the least massive companion, the age is overestimated by $\approx$\,3.5\,Gyr. This is expected:  isochrones have very little predictive power for stellar ages for such low mass stars, as they do not evolve significantly on timescales similar to the age of the Universe. Hence, the age posteriors for even the lowest mass secondaries approach our flat prior and herd age estimates towards the center of our age distribution ($\approx$\,10\,Gyr). The rest of the age--secondary mass curve behaves mostly like we expect for the 0.85\,\mdot\ star. As we add more massive secondaries, the binary system will mimic the photometry of an older, more-evolved star. The slight turnover to younger ages at the end of the curve means that the system mimics the photometry of a slightly younger but more massive primary.

For all unresolved, equal mass binary systems, we would expect to determine the same effective temperature as a single star, but a $\sqrt{2}$ times larger radius. However, we do not see this in the inset H-R diagram for the 0.85\,\mdot\ star because of the grid-edge behavior of the red ``backwards-c'' curve. Due to the finite age limit of the MIST models, the only single star models that are consistent with the higher luminosity of the binary are hotter, not purely larger, older models. The ``backwards-c'' curve also occurs because secondaries of different colors and magnitudes affect the resultant color of the system: the least massive, coolest secondaries hardly contribute to the overall photometry and the most massive secondaries are very similar in color and magnitude to the primary. Any unresolved companion $>$\,0.3\,\mdot\ will significantly affect the derived age.

The age bias is reduced for the 1.15\,\mdot\ system (yellow), which only becomes significantly affected once the secondary mass reaches half the mass of the primary. Eventually, the curve turns over as more massive secondaries push the system to larger luminosities and hence more evolved versions of higher mass stars at similar ages. In the H-R diagram, this system appears as a ``backwards-c'', mimicking the behavior of the 0.85\,\mdot\ primary system. However, the 1.15\,\mdot\ primary star system produces larger radii and similar \teff, and is unaffected by grid-edge effects when combined with similar-mass secondaries. Typical age uncertainties ($\approx$\,2\,Gyr) for these stars mean that the majority of the inferred age curve is within 1$\sigma$ of the primary star's true age except for the broad bump occurring for 0.6--1.0\,\mdot.

Finally, the blue 1.45\,\mdot\ binary age--secondary mass curve in Figure \ref{fig:binagefx} shows a slight underestimation of the ages of high mass stars. The curve increases only slightly until the secondary masses exceed 0.8\,\mdot, at which point it begins to predict even smaller ages. This slight increase in the derived system age occurs because the system's photometry appears to move slowly along its evolutionary track for companions $<$\,0.8\,\mdot. For larger secondary masses, we determine smaller ages because the combined photometry places the system on higher mass tracks where stars are both younger and at similar \teff. The behavior of this system in the H-R diagram is qualitatively different from the other systems, as the 1.45\,\mdot\ star is past the main sequence turn-off at 2.59\,Gyr. The top of the blue curve represents an equal mass binary, while evolved, smaller mass secondaries make up the top half of the ``S'' pattern (i.e. squares labeled 3 and 4). The remaining lower mass secondaries produce the bottom half of the ``S'' curve (i.e. squares labeled 1 and 2), similar to the ``backwards-c'' seen in the 0.85 and 1.15\,\mdot\ curves. Typical age uncertainties for these stars ($\approx$\,0.5\,Gyr) place the curve within uncertainties of the primary star's input age except for the most massive ($\gtrsim$\,1\,\mdot) secondaries.

Ultimately, we find that while binaries will affect our derived stellar ages, the magnitude of the effect scales with the typical error bars for each type of star. Low mass star ages are biased significantly by as many as 10\,Gyr by the presence of a mass ratio $>$\,0.5 secondary companion, but the age uncertainties we derive for these stars are on the order of 6--7\,Gyr. For higher mass stars, we find our ages are biased by only a few Gyr, where the error bars are typically on the order of 1\,Gyr.

\subsection{Accuracy of Derived Stellar Ages}\label{sec:compages}

\subsubsection{Cluster Ages}

\begin{figure}
\resizebox{\hsize}{!}{\includegraphics{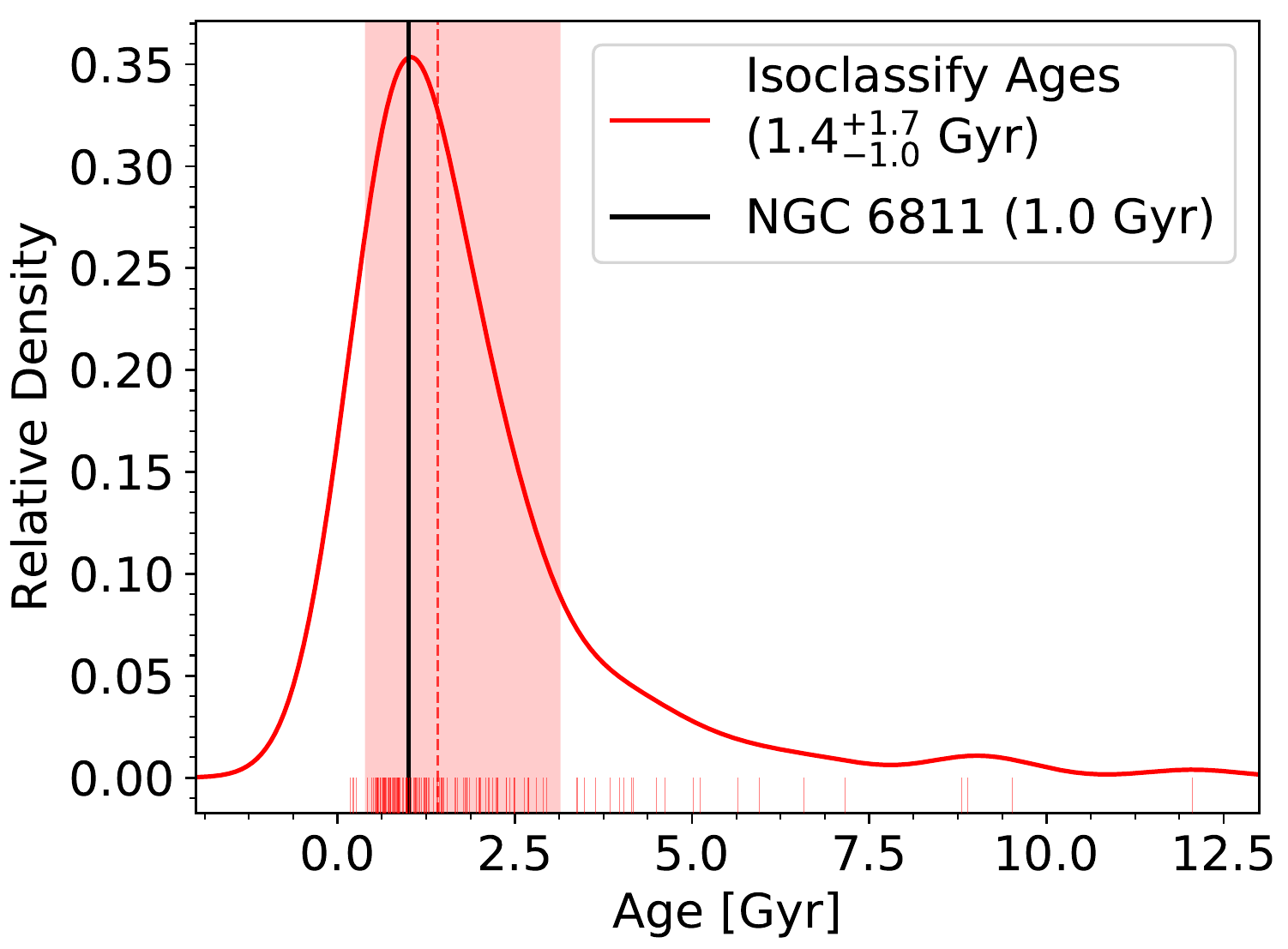}}
\caption{Age comparison for open cluster NGC 6811 at 1.0\,Gyr \citep{Meibom2011}, located within the \kep\ field. The solid red distribution represents Gaussian Kernel Density Estimate (KDE) of the ages of individual stars within each cluster as derived in this work, with the median and the 1$\sigma$ confidence interval represented by the vertical red dashed line and shaded region, respectively. We use Scott's rule \citep{Scott1992} bandwidths to produce the overall distribution. Translucent vertical red lines represent the inferred ages for each star within the sample. The black, solid vertical line represents the cluster ages from the literature in each panel. We only include non-giant stellar constituents with TAMS\,$<$\,14\,Gyr.}
\label{fig:clusteragecomp}
\end{figure}

To independently confirm our derived stellar ages, we used the 1\,Gyr open cluster NGC 6811 \citep{Meibom2011}. We did not use NGC 6791 because its main sequence turnoff was too faint for our photometry.

We utilized the \gaia\ DR2-KIC matches provided by Godoy-Rivera et al.\ (in prep) as our sample, resulting in 287 matches. We used the same output parameters as those provided in Table \ref{tab:output}. This is because the cluster's metallicity [Fe/H]\,=\,0.05\,dex \citep{Zakowicz2014} is well-within the Gaussian prior centered at solar metallicity (0.0\,dex) with the standard deviation of 0.2\,dex that we assume for \kep\ field stars without spectroscopic metallicities. To ensure that our NGC 6811 ages are reliable, we removed all stars with TAMS\,$>$\,14\,Gyr and all giant stars using an ad-hoc cut in stellar radius-\teff\ space, similar to that of Equation (1) in \cite{Fulton2017}: $\frac{R_\star}{\mathrm{R_\odot}} < 10^{0.00035(T_{\mathrm{eff}}-4500) + 0.15}$. In addition, we removed all stars with RUWE\,$>$\,1.2 to minimize potential age-contaminating binaries (see \S\ref{sec:binagefx}). This left us with 146 matches in NGC 6811.

NGC 6811 shows good agreement between the literature cluster age and the isochrone-dependent ages we derive here (Figure \ref{fig:clusteragecomp}). There are a few stars in the high-age tail of the distribution, but these are stars that populate the coolest part of the remaining main sequence (TAMS\,$\lesssim$\,14\,Gyr) and have the largest radii. Although we already removed all stars with high RUWE, these targets might be close or unresolved binaries with a hotter and cooler component, which biases the resulting photometry to indicate a cooler star of a larger radius and hence an older age. This effect can be large, as discussed in \S\ref{sec:binagefx} and demonstrated in Figure \ref{fig:binagefx}. However, only a few stars lie in such areas of parameter space.

\subsubsection{Asteroseismic Ages}

\begin{figure}
\resizebox{\hsize}{!}{\includegraphics{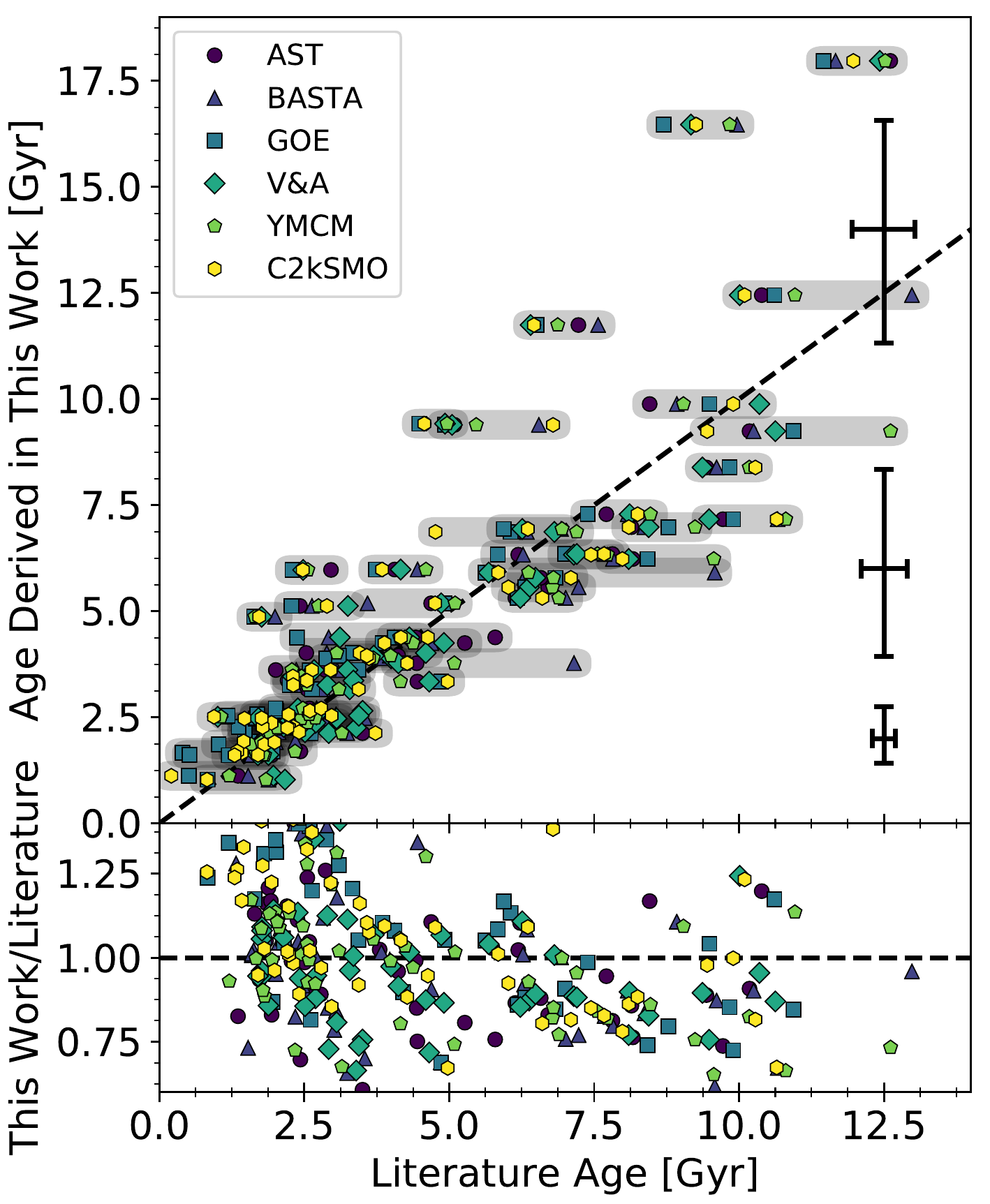}}
\caption{Ages derived in this work versus those with frequency-modeled asteroseismic ages from a variety of pipelines \citep{silva17}. The black dashed line represents agreement. The various colors/shapes represent the different pipelines. The translucent grey rounded rectangles represent the ranges of age estimates from different asteroseismic pipelines for each system, which includes not only seismic differences but also differences between model grids. The bottom panel is the ratio of the two age determinations. We have also plotted median error bars in the right-hand portion of the top panel, where, from bottom to top, the error bars represent the median uncertainties of stars with isochrone ages between 0--4\,Gyr, 4--8\,Gyr, and $>$\,8 Gyr, respectively.} 
\label{fig:astagecomp}
\end{figure}

We also compared our derived ages to those of \kep\ stars that have asteroseismic ages. We utilized the ``boutique'' frequency-modeled ages from the \kep\ legacy sample detailed in \cite{silva17}, which includes results from a number of analysis pipelines. In Figure \ref{fig:astagecomp}, the ages that we derive are in reasonable agreement with those provided by a variety of asteroseismic pipelines. In the top panel of Figure \ref{fig:astagecomp}, we plot each asteroseismic pipeline as a different color, so horizontal rows of colored points indicate results for the same star. The horizontal scatter of the colored points are typically larger than their reported errors, which indicates that systematic pipeline differences dominate. Any deviations from the 1:1 dashed line are sufficiently accounted for by a combination of the typical error bars (bottom right, top panel) and any systematic scatter depending on the asteroseismic pipeline one chooses. In addition, the asteroseismic ages do not fall above the age of the Universe (likely due to a model grid age-cutoff), hence we see the largest differences at older ages. However, even the largest discrepancies are within 1$\sigma$ of the 1:1 line.

Ultimately, we report a median offset and scatter of 5\% and 29\%, respectively. We conclude that our isochrone-derived ages are consistent with ages determined through more precise methods within the uncertainties that we report.

\subsubsection{Kinematic Ages}

\begin{figure}
\centering
\resizebox{\hsize}{!}{\includegraphics{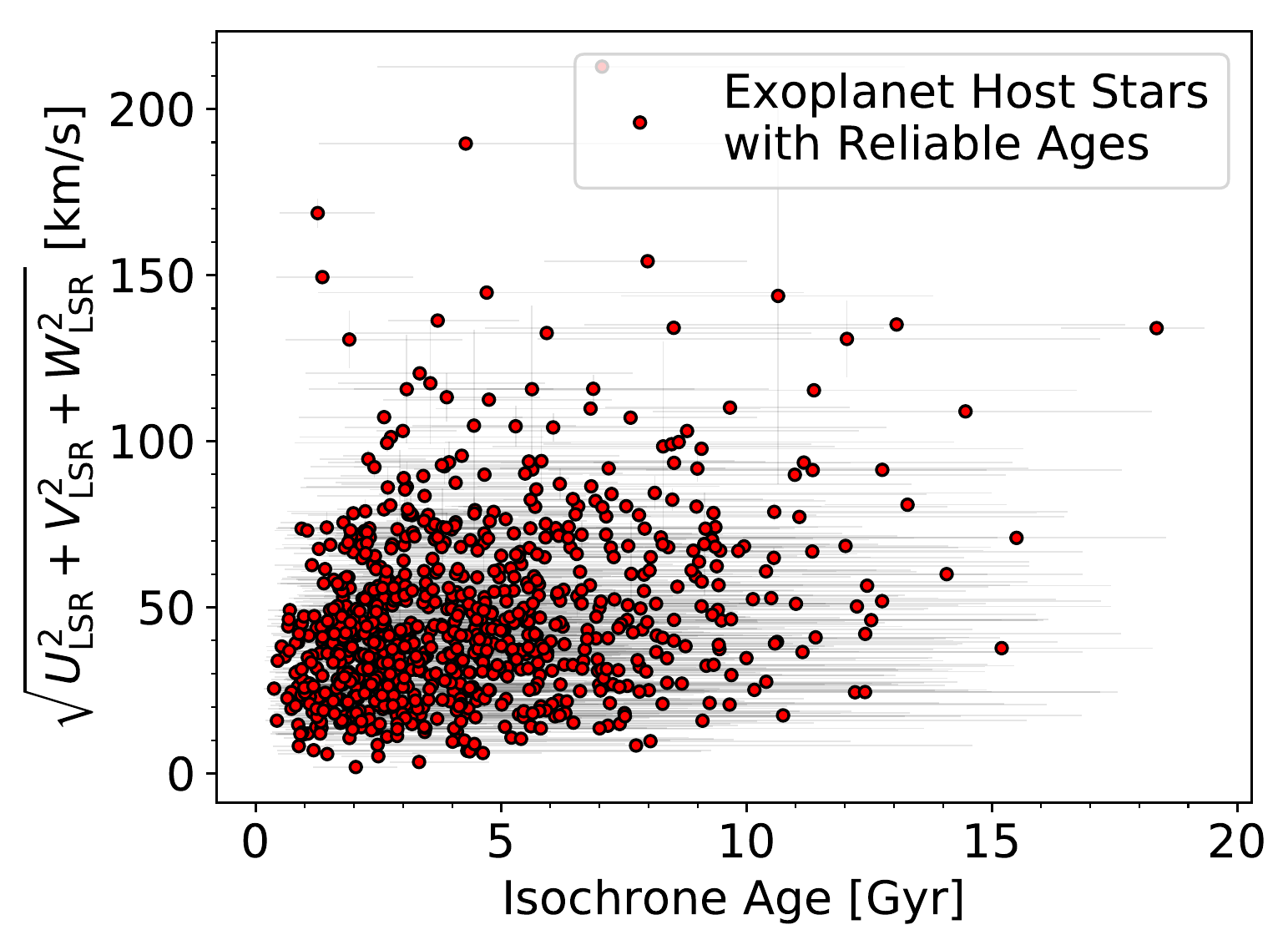}}
\caption{Total space (UVW) velocities relative to the local standard of rest derived from \gaia\ DR2 proper motions and parallaxes and CKS \citep{Petigura2017} radial velocities versus isochrone ages computed in this work for \kep\ exoplanet host stars with reliable ages. In this case, host stars with reliable ages are dwarfs with spectroscopic metallicities, RUWE\,$<$\,1.2, TAMS\,$<$\,20\,Gyr, and \texttt{iso\_gof}\,$>$\,0.99. We plot uncertainties as grey, translucent error bars.}
\label{fig:kinage}
\end{figure}

In Figure \ref{fig:kinage} we display a comparison of our isochrone-derived ages for \kep\ exoplanet host stars with spectroscopic metalllicities, RUWE\,$<$\,1.2, TAMS\,$<$\,20\,Gyr, and \texttt{iso\_gof}\,$>$\,0.99 with their total space (UVW) velocities relative to the local standard of rest (LSR). We computed the latter from \gaia\ DR2 proper motions and parallaxes and CKS \citep{Petigura2017} radial velocities, following the method outlined in \cite{Newton2016}:  we used equation (1) from \cite{johnson87} and the transformation matrix from \cite{perryman97} with the LSR defined by \cite{Schonrich2010}. We computed uncertainties from equation (2) in \cite{johnson87}, where we used 0.1\,km/s radial velocity uncertainties \citep{Petigura2017} and the formal \gaia\ DR2 uncertainties on proper motions and parallaxes \citep{Brown2018,Lindegren2018}. Most total space velocity uncertainties are smaller than the markers. Figure \ref{fig:kinage} reveals that both the mean total space velocities and their dispersion increase at higher isochrone ages, which matches the expectations that old stars have had more time to be perturbed by gravitational interactions with other stars \citep{Soderblom2010,Newton2016}.

\section{Revised Properties of \kep\ stars}\label{sec:revstars}

\begin{figure*}
\resizebox{\hsize}{!}{\includegraphics{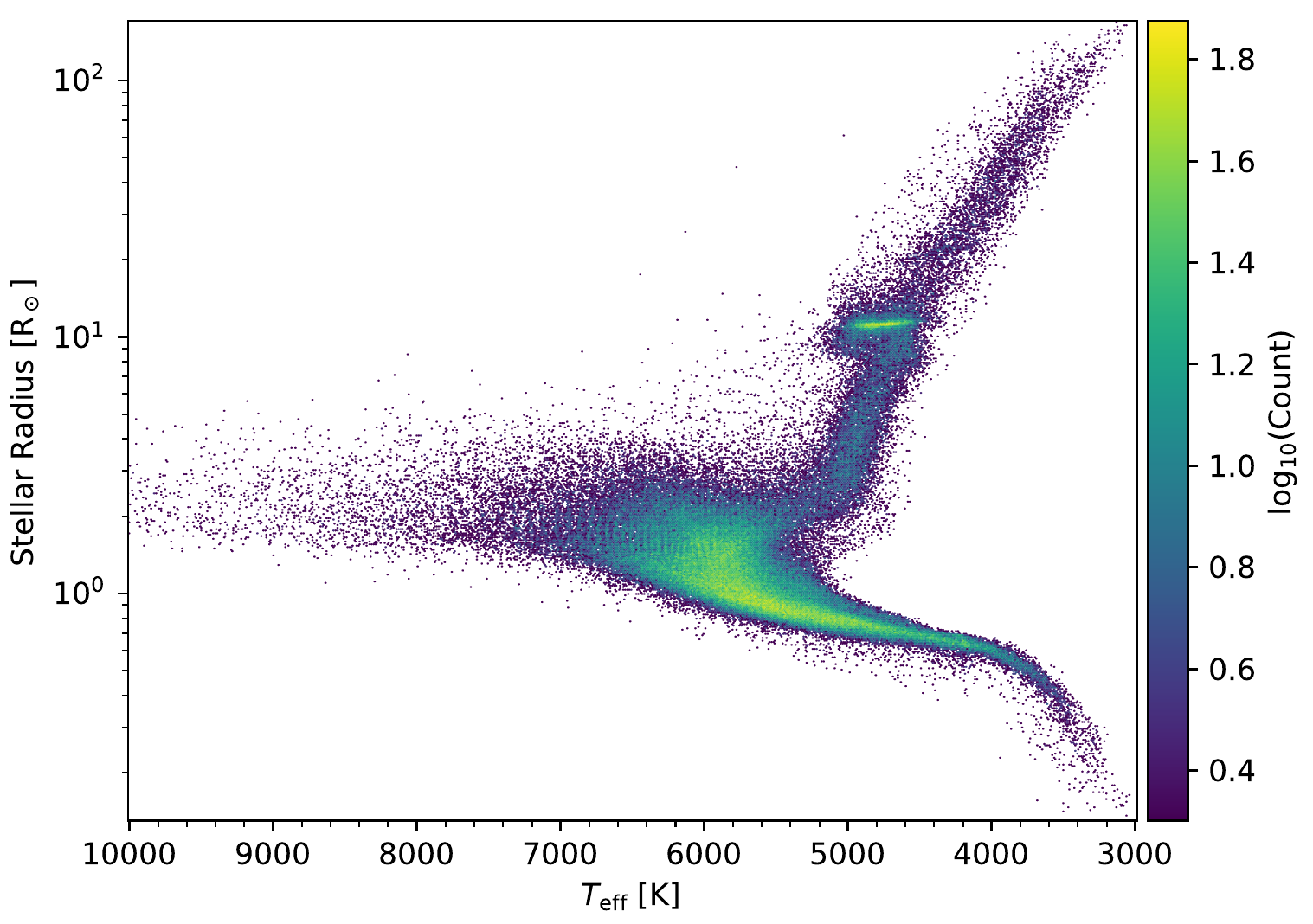}}
\caption{Radius versus effective temperature for $\sim$\,186,000 \kep\ stars with radii and \teff\ derived based on $Gaia$ DR2 parallaxes and $g - K$ photometry presented above. Color-coding represents logarithmic number density.} 
\label{fig:HR}
\end{figure*}

\subsection{Catalog Description}\label{sec:catdescrip}

Here, we investigate the properties of all \nstars\ \kep\ stars. We tabulate stellar \teff, \logg, metallicity, mass, radius, luminosity, mean stellar density, age, distance, and extinction in Table \ref{tab:output}. Some stars in Table \ref{tab:input} are not included in Table \ref{tab:output}. These stars have fewer than ten models within 4$\sigma$ of their observational uncertainties, so their under-sampled posteriors are insufficiently constrained. Stars without a solution frequently appear in unphysical areas of parameter space for a single star, such as above the lower main sequence. In addition, we flagged the ages (and corresponding uncertainties) of stars which we deem unreliable or uninformative with asterisks. Unreliable ages are flagged according to the goodness-of-fit parameter (GOF), which is computed using the overall likelihood of the closest model grid point to the set of input observables found in Table \ref{tab:input}. We provide the GOF parameter in Table \ref{tab:output}; we recommend treating any GOF\,$<$\,0.99 stars with extra caution. Stars with GOF\,$<$\,0.99 are outliers in stellar radius--\teff\ space and they have extremely small fractional error bars compared to typical stars within our catalog. We chose 0.99 as our GOF cut because it represents a compromise between keeping too many outliers (GOF\,$<$\,0.9) and removing too many stars with reasonable solutions (GOF\,$<$\,0.999) based on the density of our computed grid.

We flag uninformative ages based on the terminal age of the main sequence (TAMS) for that star. We compute the TAMS by performing 2D interpolation on MIST evolutionary tracks of stars of similar mass to each of our derived stellar masses. If the TAMS of the star is greater than the maximum age of our grid, 20\,Gyr, we do not expect to derive any informative age information from that star, given the observational uncertainties and the limitations of isochrone placement. We choose to use 20\,Gyr as our age cutoff rather than the age of the universe because we still determine informative, non-truncated age posteriors for stars older than $\approx$\,14\,Gyr from our 20\,Gyr maximum age grid. About 14\% of stars within the \kep\ field have TAMS\,$>$\,20\,Gyr. Therefore, most K and all M-dwarfs have uninformative ages, as these stars have not had enough time within the age of the universe to evolve substantially in the H-R diagram.

\subsection{The Grid-Modeled H-R Diagram of \kep\ Stars} \label{sec:HR}

Figure \ref{fig:HR} shows stellar radius versus effective temperature for the \kep\ stars with grid-modeled radii and \teff\ determined by this work. We see a clear main sequence, from M dwarfs at \teff\,=\,3000\,K and $R_\star$\,$\approx$\,0.2\,$R_\odot$, through A stars at \teff\,$\lesssim$\,9000\,K and $R_\star$\,$\approx$\,2\,$R_\odot$. The main sequence turnoff at \teff\,$\approx$\,6000\,K and $R_\star$\,$\approx$\,2\,$R_\odot$ is visible, along with the giant branch. We identify the ``red clump'' as the concentration of stars surrounding \teff\,$\approx$\,4900\,K and $R_\star$\,$\approx$\,11\,$R_\odot$. As expected, the \kep\ catalog is dominated by F and G-type stars as a result of the selection bias for solar-type stars to detect Earth-like transiting exoplanets \citep{batalha10}.

Unlike all previous \kep\ Stellar Properties Catalogs \citep{huber14,Mathur2017,Berger2018c}, the \teff\ gap around 4000\,K is no longer present. We expect this given that we are deriving \teff\ from our continuous set of input $g-K_s$ colors and the \cite{mann15} color-\teff\ relation overlaps at 4200\,K, as we discussed in \S\ref{sec:inputphotometry} and \S\ref{sec:teff}. Stellar input observables and their uncertainties must include at least ten MIST models to produce a solution, so we do not find any stellar solutions outside of our model grid. Therefore, we do not report output parameters for 247 stars in Table \ref{tab:output}. While some stars fall below the nominal main sequence, the discrepancies are not as large as those reported by \cite{Berger2018c}. A number of stars ($\sim$\,500) below the main sequence that are inferred to be subdwarfs (\teff\,=\,3600--5400\,K and $R_\star$\,$<$\,0.6\,\rdot) or in other extreme parameter regimes could have erroneous \teff\ values or excess noise in the astrometry according to \gaia\ DR2. In addition, if we ignore all stars with RUWE\,$>$\,1.4 (reducing \nstars\ to $\approx$\,170,000 stars), the putative subdwarfs disappear, as well as a number of other stars in sparsely-populated areas of the H-R diagram. Therefore, most of the inferred subdwarfs have high RUWE values and thus potentially erroneous parallaxes.

The binary main sequence identified in \cite{Berger2018c} is not prominent here. This has two reasons: (1) we have corrected photometry for stars with secondaries between 1 and 4'' (\S\ref{sec:binaries}) and (2) lower main sequence grid models, even out to ages of 20\,Gyr, are still not luminous (and hence large) enough to emulate the absolute magnitude of a multiple-star system -- thus, such results are not allowed by our analysis.

The striping pattern ranging from 5800--7000\,K and $\approx$\,1--2\,\rdot\ is an artifact of our model grid. The brighter colored stripes are stellar isochrones at solar metallicity, where individual stellar solutions preferentially ``snap'' to the isochrone grid. Increasing the age resolution of our model grid or only using stars with spectroscopic metallicity constraints would significantly reduce the contrast of the stripes, but computational constraints and the lack of spectroscopic metallicities for 2/3 of the \kep\ sample prevent us from doing so here.

\subsection{The Grid-Modeled Mass-Luminosity Diagram of \kep\ Stars} \label{sec:LumMass}

\begin{figure*}
\resizebox{\hsize}{!}{\includegraphics{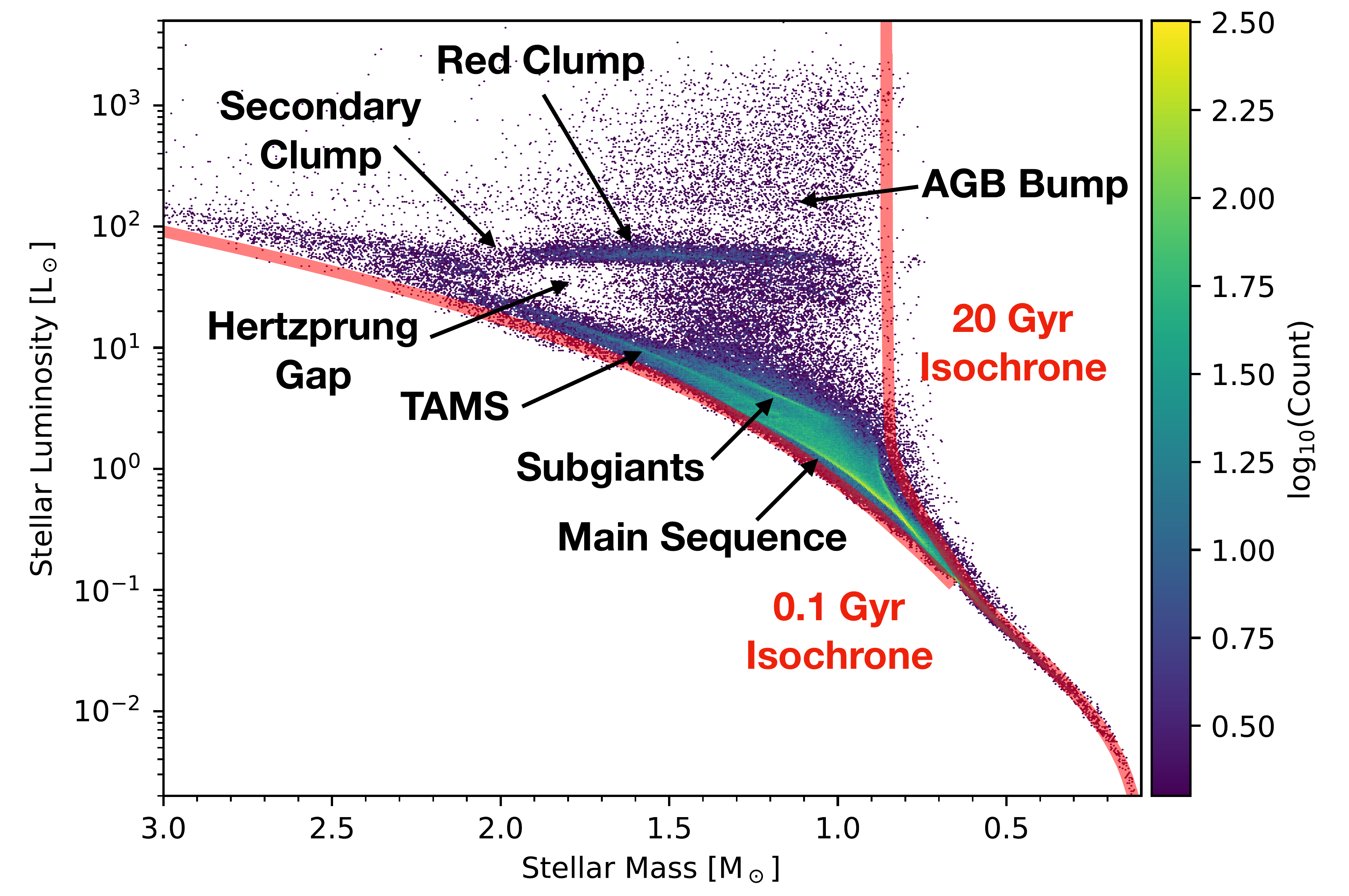}}
\caption{Luminosity versus mass for $\sim$\,186,000 \kep\ stars. Color-coding represents logarithmic number density. The red, translucent curves represent the 0.1 (left) and 20\,Gyr (right), [Fe/H]\,=\,0.0\,dex isochrones. We have labeled all features in the distribution accordingly.} 
\label{fig:LumMass}
\end{figure*}

Another benefit of isochrone fitting over the work in \cite{Berger2018c} is the determination of stellar masses. In Figure \ref{fig:LumMass}, we plot our grid-modeled luminosities versus our grid-modeled masses. This diagram shows a variety of features resulting from the processes of stellar evolution, similar to the radius-\teff\ H-R diagram in Figure \ref{fig:HR}. It is even easier to see the stellar radius evolution in this plot, given the ability to choose a mass on the x-axis and follow the change in density of points as the stellar age (and luminosity) increases.

For masses below 0.8\,\mdot, the main sequence grows thinner due to the lack of luminosity evolution within the age of the universe. We see that for stellar masses between 0.6 and 0.8\,\mdot, there is some scatter around the ZAMS. This is close to where we replaced the MIST model parameters with the M-dwarf empirical relations from \cite{mann15} and \cite{Mann2019} (\S\ref{sec:modelgrid}). From there, the luminosities drop off as well as the apparent scatter as the masses approach the hydrogen burning limit.

At masses $>$\,0.8\,\mdot, we see the distribution expands vertically. In addition, the highest density of points occurs between 0.8 and 1.0\,\mdot, representing the large fraction of solar-type stars within the \kep\ sample. For masses $\gtrsim$\,0.9\,\mdot, luminosities begin to span from the main sequence up the giant branch. The smooth curves tracing the outermost models on both the left and the right represent the minimum and maximum age solar-metallicity isochrones that we used in our analysis. The lower-left edge is a 100\,Myr isochrone, while the upper right edge is a 20\,Gyr isochrone.

There are a few features that are prominent as a function of mass and luminosity at masses $\gtrsim$\,1.0\,\mdot. Starting from the bottom of the distribution, we see that there is an over-density of points arcing from the yellow, highest densities up and to the left to subsequently higher masses and luminosities. This branch represents some of the youngest stars in our sample, less than half their TAMS. Just above this main sequence curve is the higher mass TAMS, the long, arcing over-density of points from $M_\star$\,$\approx$\,1.1--3.0\,\mdot\ and $L_\star$\,$\approx$\,2.5--140\,\ldot. Next, we see an over-density of points ranging from $M_\star$\,$\approx$\,1--1.7\,\mdot\ and $L_\star$\,$\approx$\,3--20\,\ldot. These stars are all subgiants, where luminosities stay roughly constant at a particular stellar mass as they move towards the red giant branch.

As we increase in stellar luminosity, we see a lack of stars at $M_\star$\,$\approx$\,1.6--2.0\,\mdot\ and $L_\star$\,$\approx$\,15--50\,\ldot, an illustration of the Hertzprung Gap. This under-density occurs because these massive stars evolve so quickly during their subgiant and giant phases that they reach the red clump almost instantaneously. The red clump is the swath of points from just above the Hertzprung Gap ($M_\star$\,$\approx$\,2.0\,\mdot\ and $L_\star$\,$\approx$\,60\,\ldot) directly to the right and lower masses at the same luminosity ($M_\star$\,$\approx$\,1.0\,\mdot\ and $L_\star$\,$\approx$\,60\,\ldot). The secondary clump (for massive stars) is the clustering of points below and to the left of the main clump for $M_\star$\,$\approx$\,2.0\,\mdot. Finally, the slight over-density of points at $M_\star$\,$\approx$\,1.1--2.0\,\mdot\ and $L_\star$\,$\approx$\,140\,\ldot\ represents the asymptotic giant branch (AGB) bump, where a few stars have reached the AGB phase. The AGB bump is analagous to the AGB as the RGB bump is to the RGB, where the luminosity decreases after the He-core and H-shell burning stops for a short time, which then causes contraction and subsequent reignition. The star then continues to grow more luminous as it moves up the AGB. This temporary deceleration of evolution on the AGB produces the AGB bump \citep{Gallart1998}. 

We cannot glean additional structure from luminosities $>$\,140\,\ldot, but we do see a wide range of masses at these high luminosities as expected. Moreover, we find that the masses of \kep\ stars range between 0.1 and 5.1\,\mdot, and only $\sim$\,400 \kep\ stars have masses exceeding 3.0\,\mdot. This is consistent with the inferred lack of very young stars in the \kep\ field.

\subsection{Parameter and Uncertainty Distributions}
\label{sec:dists}

\subsubsection{Temperatures, Surface Gravities, and Metallicities}
\label{sec:spechist}

\begin{figure*}
\includegraphics[width=0.50\textwidth]{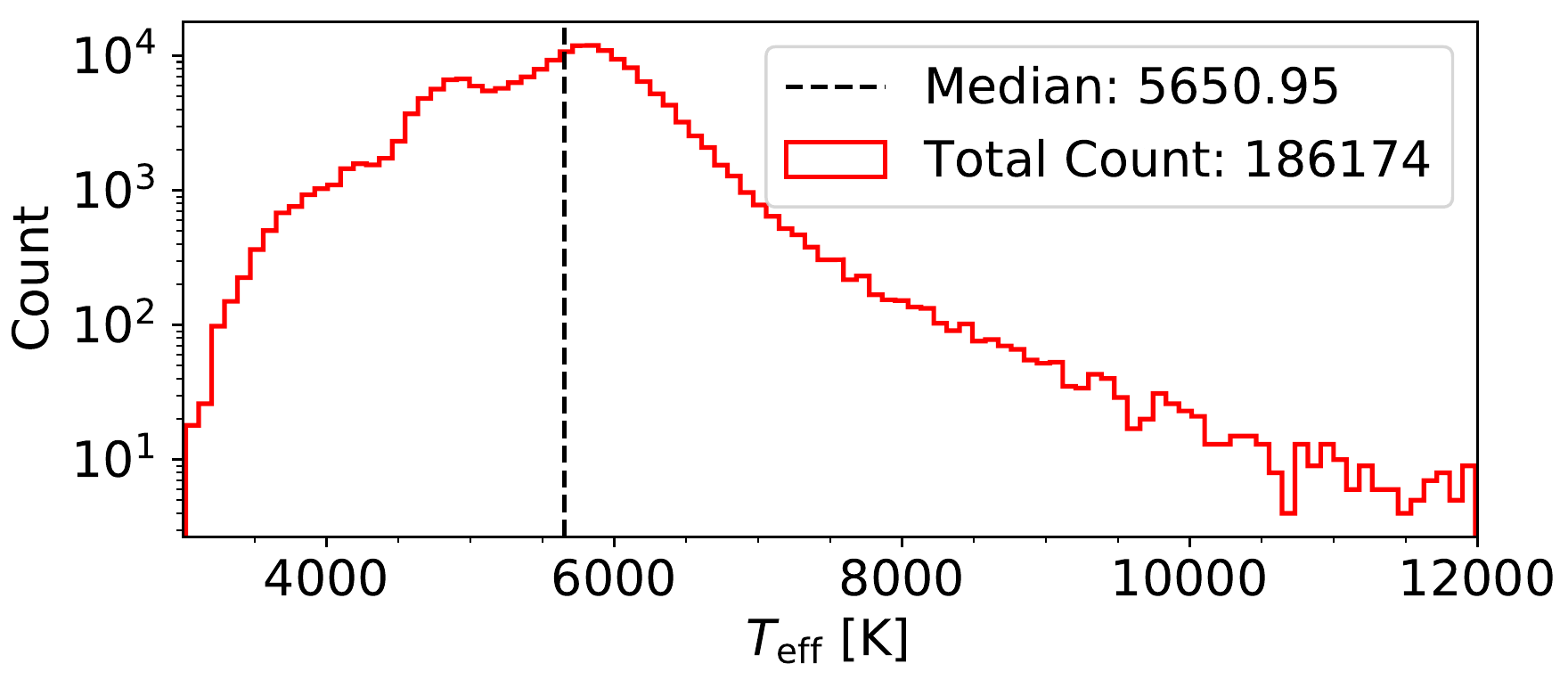}
\includegraphics[width=0.50\textwidth]{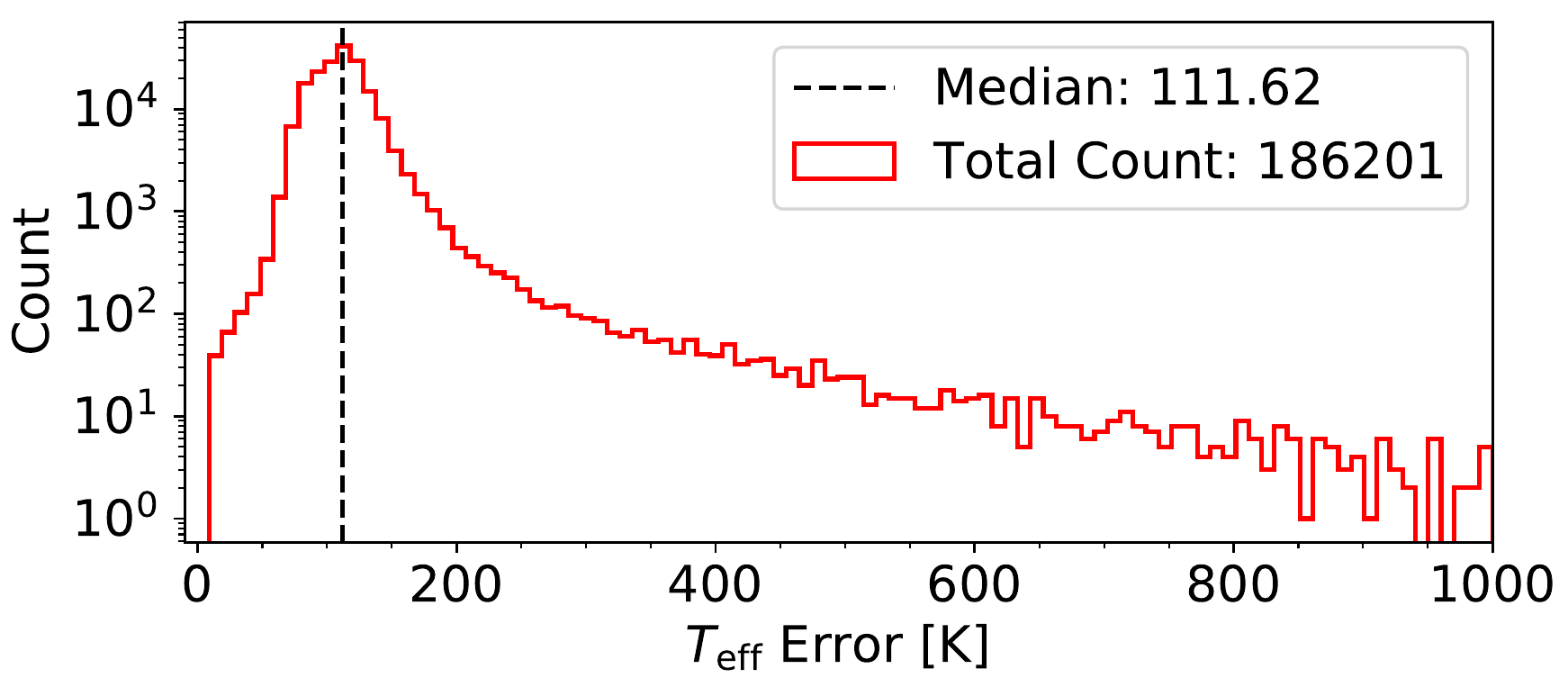}
\includegraphics[width=0.50\textwidth]{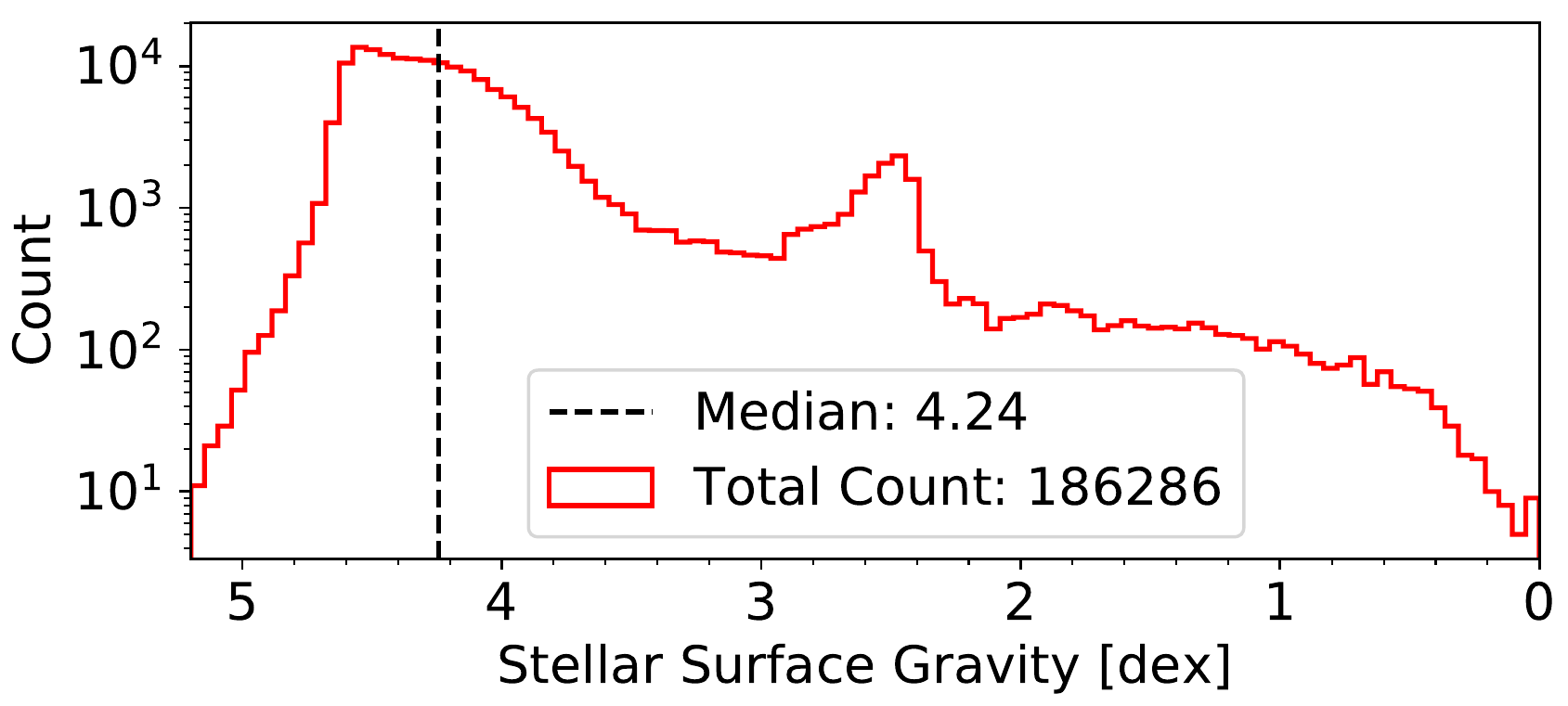}
\includegraphics[width=0.50\textwidth]{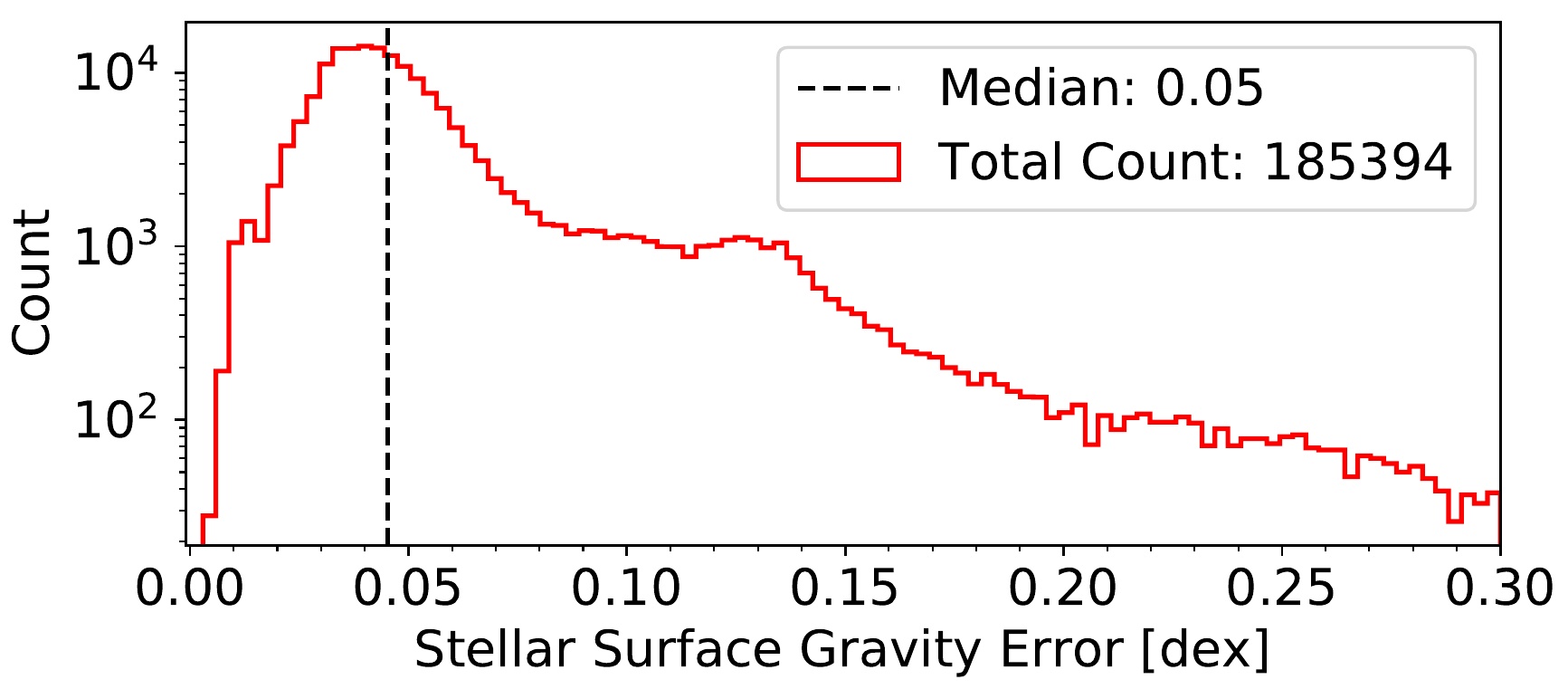}
\includegraphics[width=0.50\textwidth]{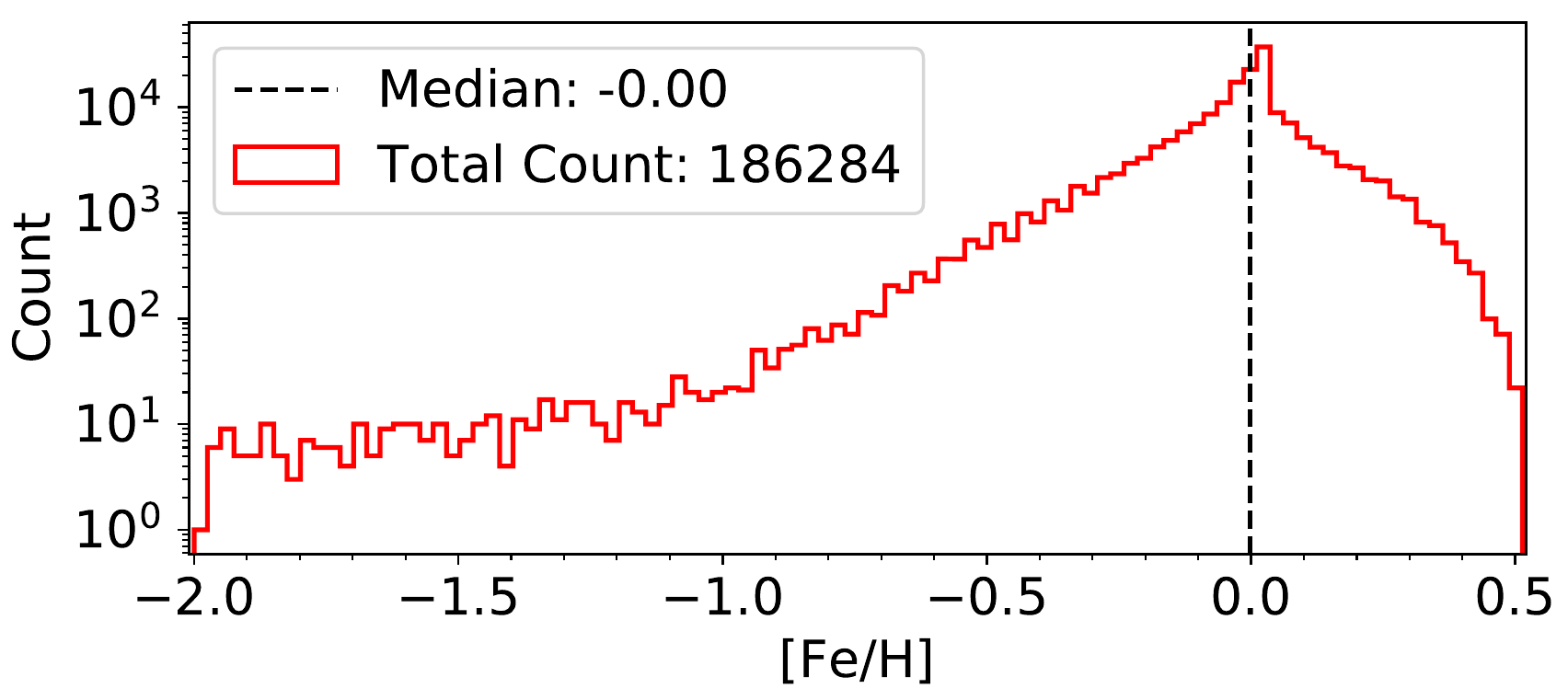}
\includegraphics[width=0.50\textwidth]{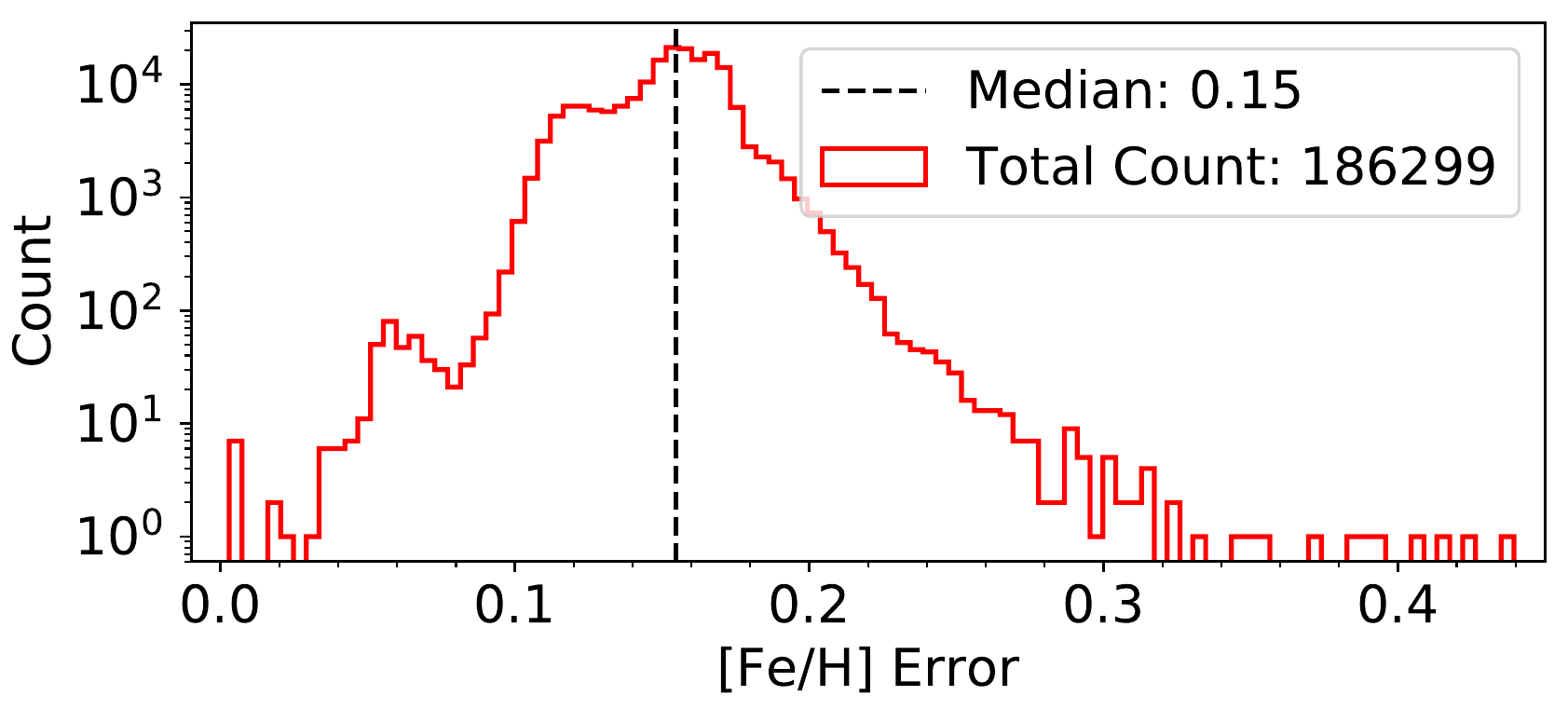}
\caption{\teff, \logg, and metallicity parameter and uncertainty distributions from our catalog. The black dashed vertical lines illustrate the median value for each parameter, the value of which is given in the legend belonging to each plot. In addition, the total number of stars in displayed in each histogram is provided in the legend. These numbers vary due to choices in parameter cutoffs, and they are usually smaller than the total number of stars presented here, \nstars. Some histograms have logarithmic scaling on the x and/or y-axes. $Top\ Row$: Stellar effective temperatures and their absolute uncertainties. $Middle\ Row$: Stellar surface gravities and their absolute uncertainties. $Bottom\ Row$: Stellar metallicities and their absolute uncertainties.} 
\label{fig:SpecHist}
\end{figure*}

In Figure \ref{fig:SpecHist}, we plot the histograms of \teff, \logg, and metallicity for the \kep\ parent sample. Unsurprisingly, the \teff\ histogram peaks close to the solar \teff, and the giants cause the slight bump around 4800\,K. The errors in \teff\ peak strongly around the median of 112\,K by design (see \S\ref{sec:modelgrid} for details), as this represents a fractional error of $\sim$\,2\% for the median star in our sample.

The peak in the stellar surface gravity (\logg) histogram occurs at 4.24\,dex in cgs units. This is consistent with the larger-than-expected percentage of \kep\ subgiant targets \citep[21\%,][]{Berger2018c}. Typical errors are on the order of 0.05\,dex, which is a dramatic improvement over the $\approx$\,0.2\,dex \logg\ median error provided in \cite{Mathur2017} due to the strong radius constraints from \gaia\ DR2 parallaxes.

Because the vast majority of \kep\ stars do not have spectroscopic metallicities ($\approx$\,120,000 out of $\approx$\,186,000), the metallicity distribution and its uncertainties are not particularly informative. The left plot shows a peak at solar metallicity with a sharp drop-off to either higher or lower metallicities. This is unsurprising, given that our priors are centered on solar metallicity. The uncertainty distribution has one major peak around the median uncertainty of 0.15\,dex; this peak represents a convolution of metallicities derived from a Gaussian prior centered at solar metallicity with a $\approx$\,0.2\,dex width and the remaining objects that have spectroscopic metallicities with fixed 0.15\,dex uncertainties. Almost all of the $\approx$\,400 stars with metallicity uncertainties $<$0.08\,dex have GOF\,$<$\,0.99 and are therefore unreliable. The remaining low-metallicity uncertainty stars have observables which place them in sparse areas of the model grid.

\subsubsection{Radii, Masses, Densities, Luminosities, and Ages}
\label{sec:evhist}

\begin{figure*}
\includegraphics[width=0.50\textwidth]{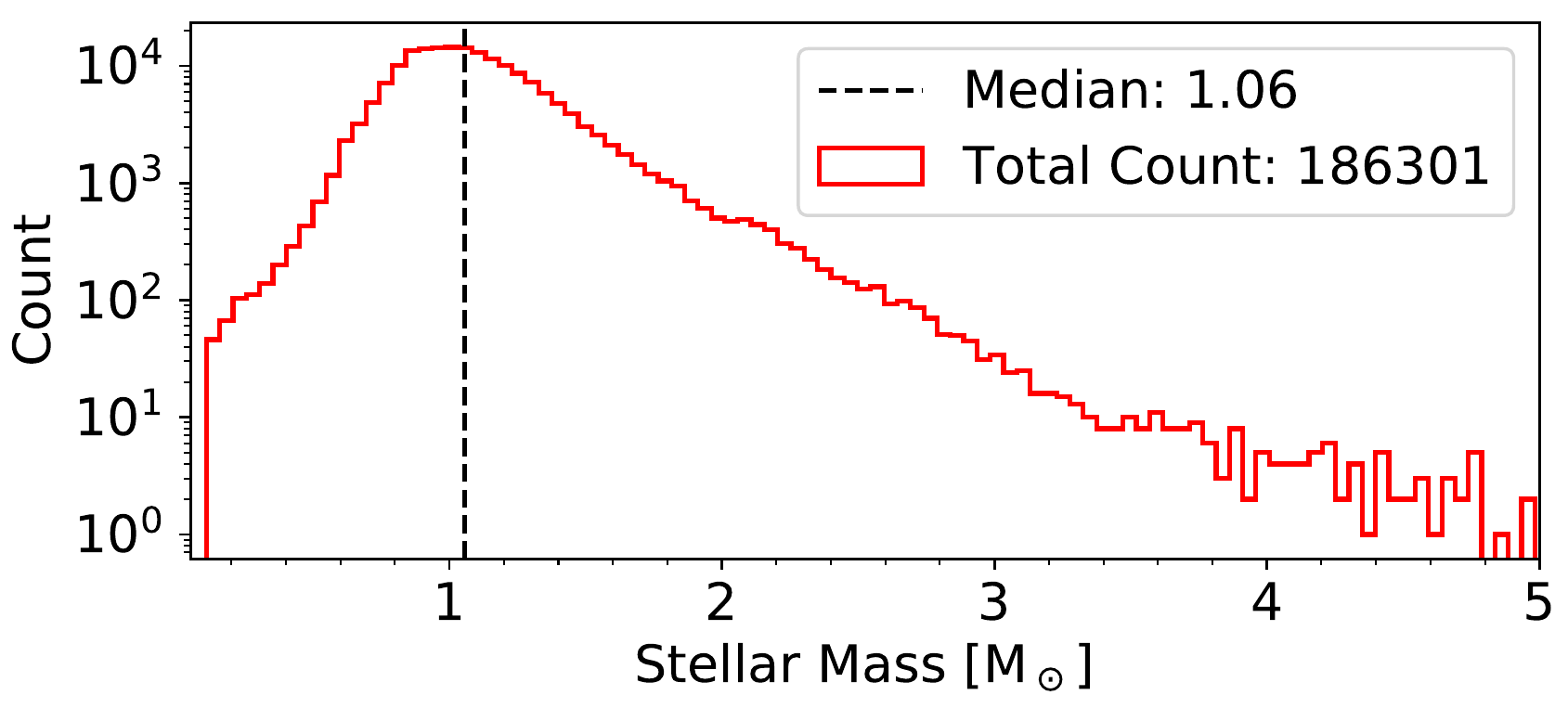}
\includegraphics[width=0.50\textwidth]{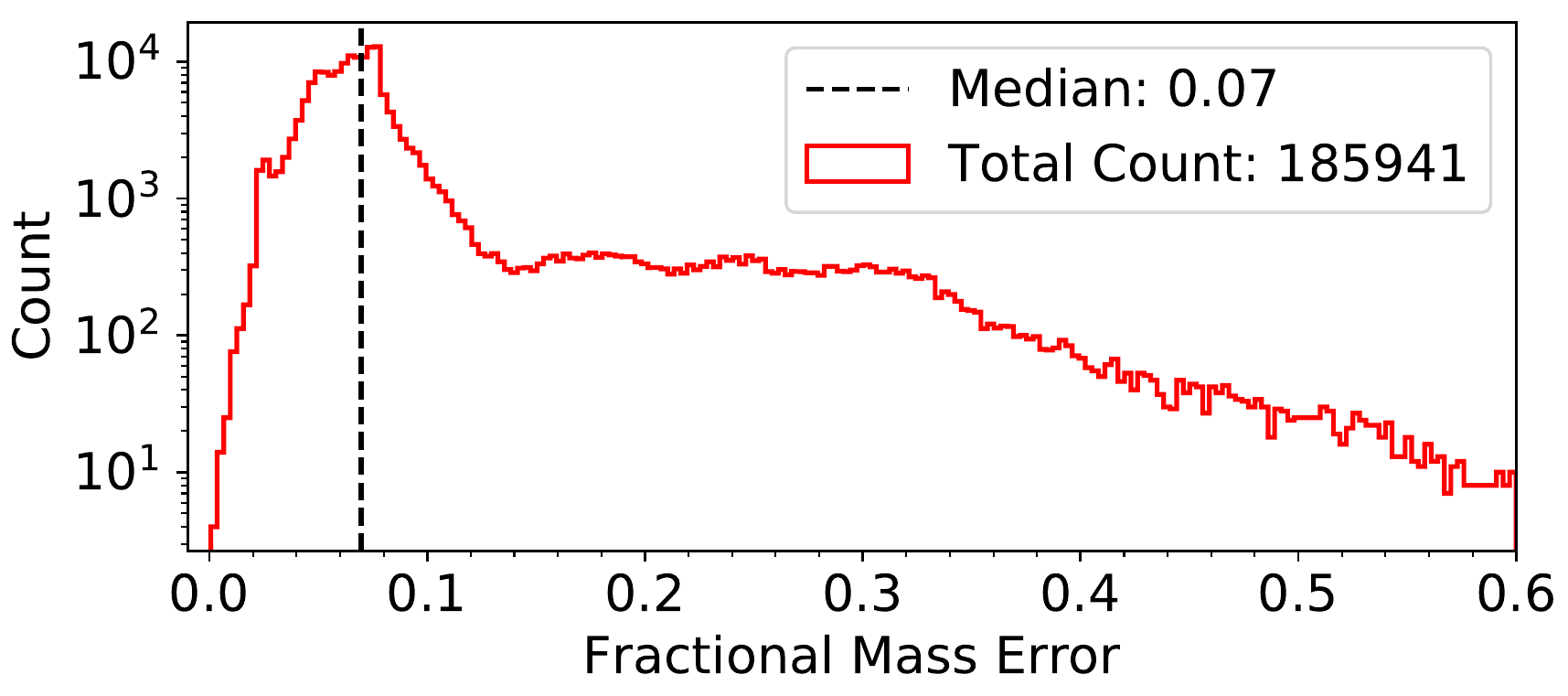}
\includegraphics[width=0.50\textwidth]{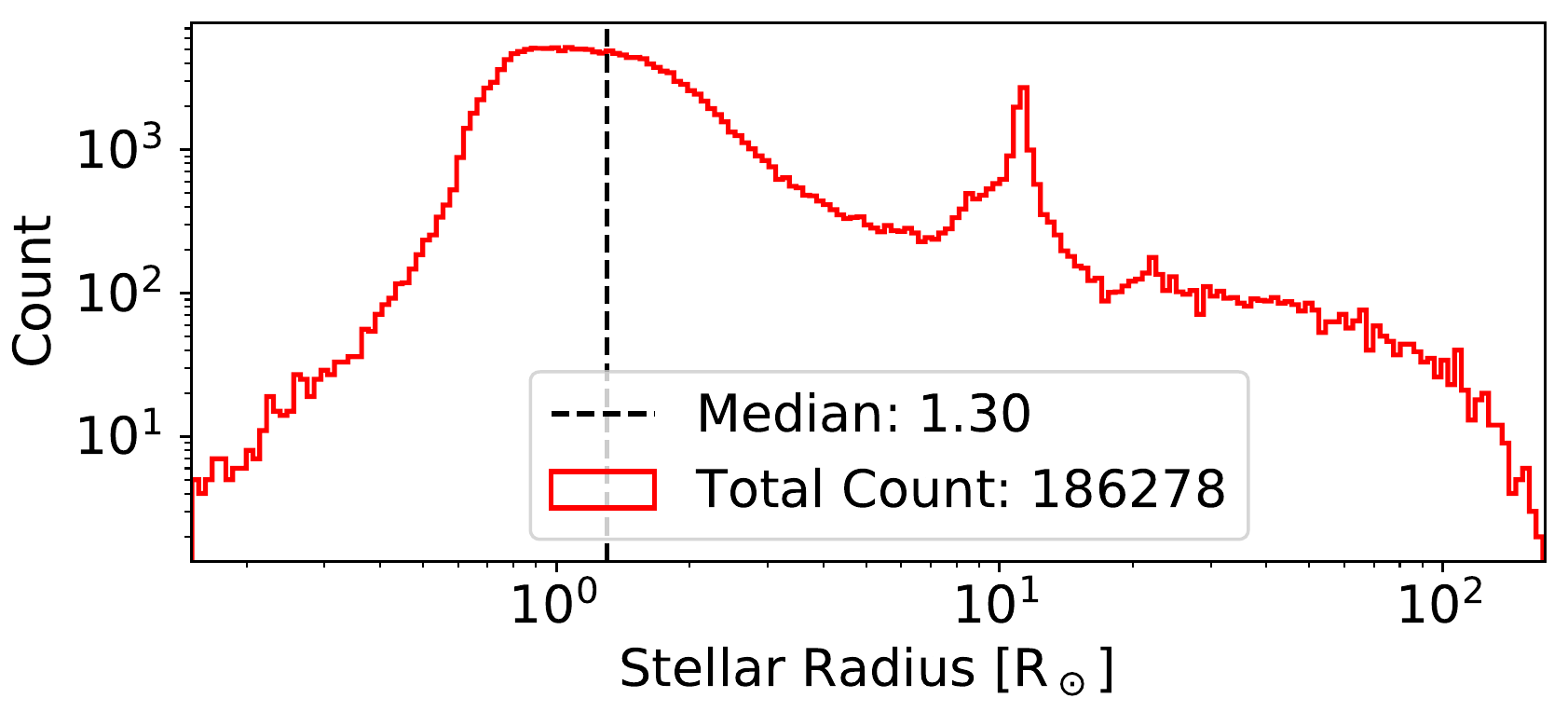}
\includegraphics[width=0.50\textwidth]{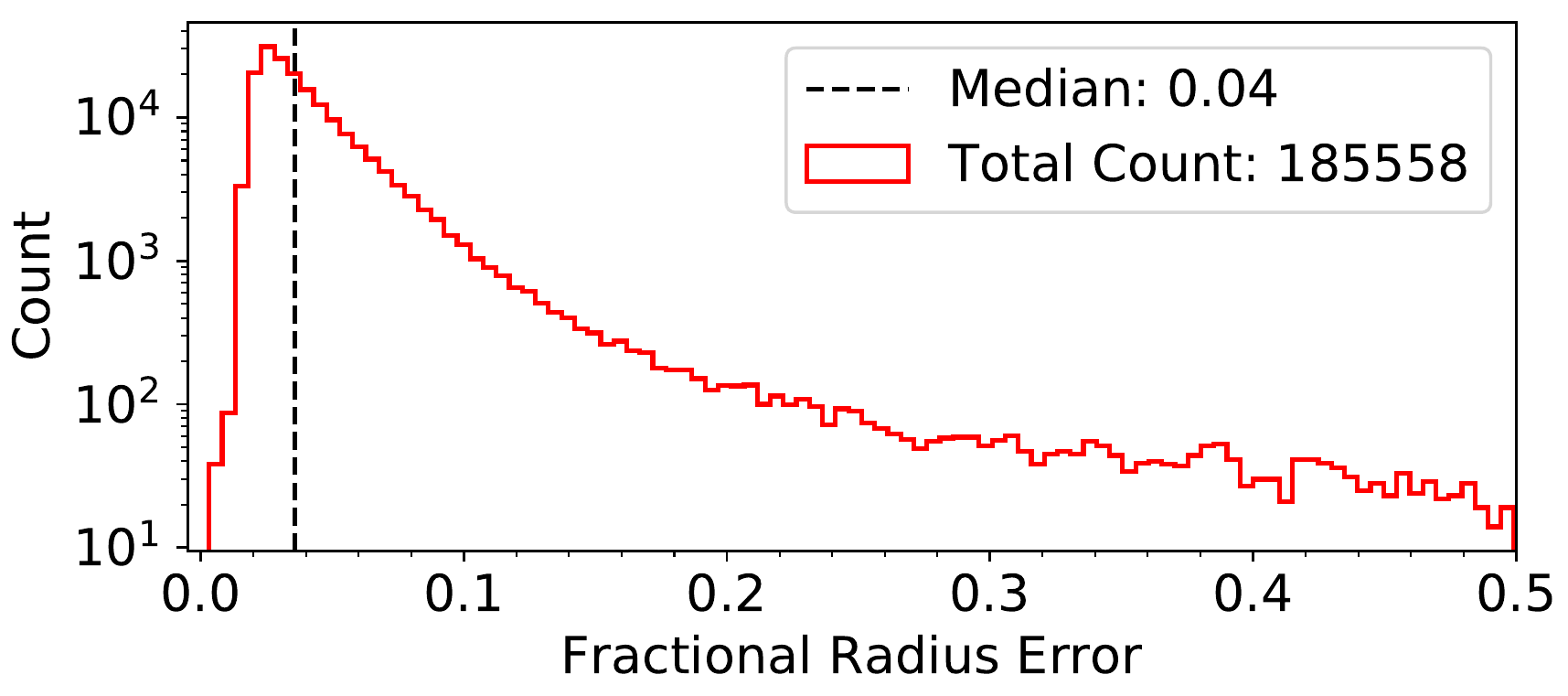}
\includegraphics[width=0.50\textwidth]{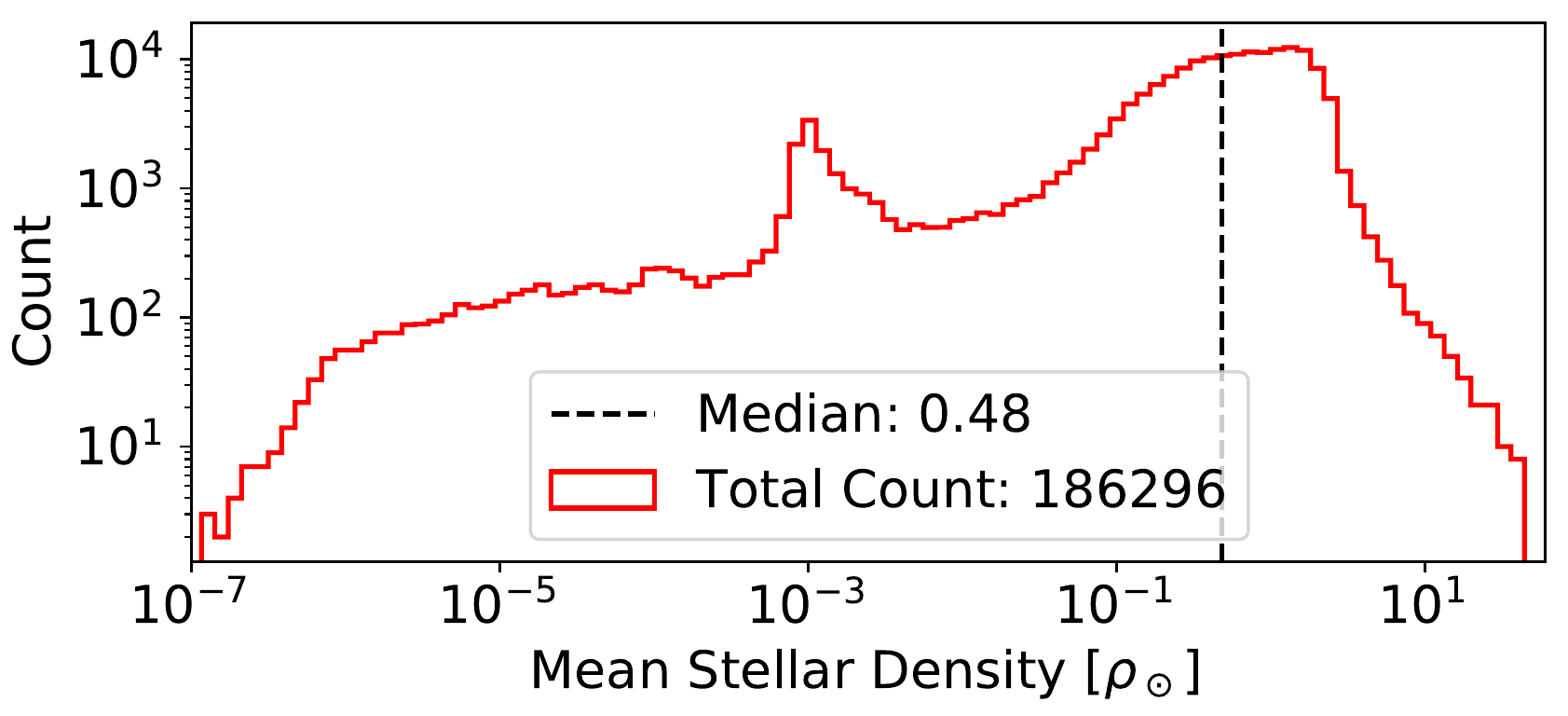}
\includegraphics[width=0.50\textwidth]{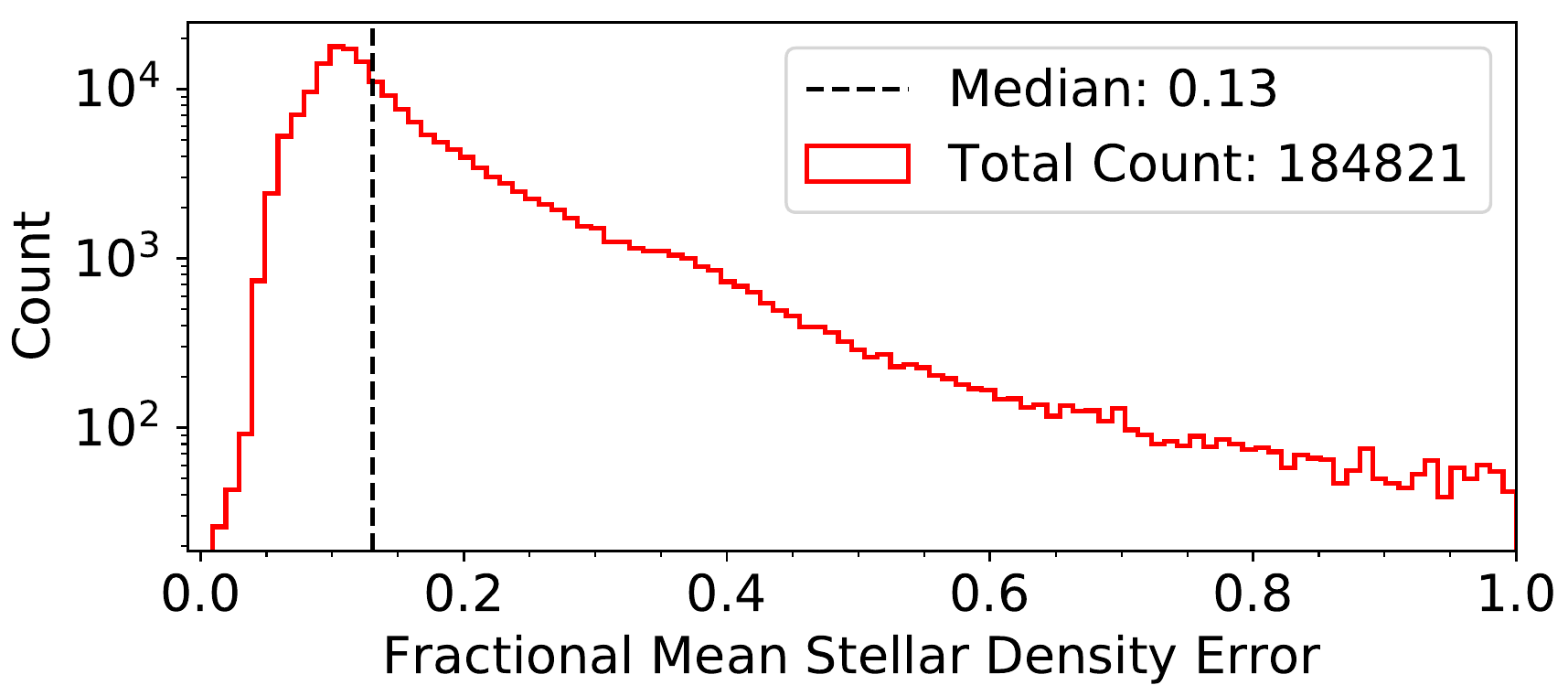}
\includegraphics[width=0.50\textwidth]{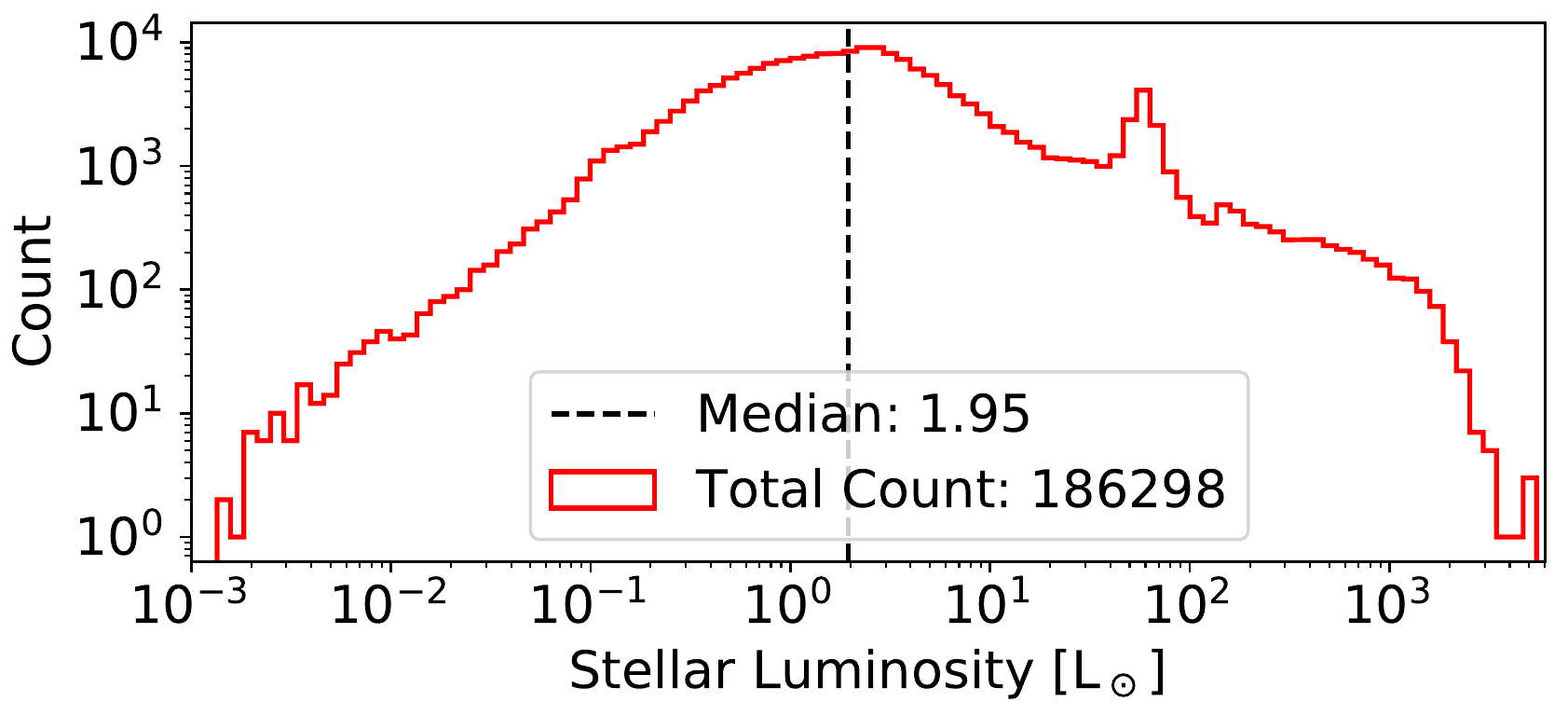}
\includegraphics[width=0.50\textwidth]{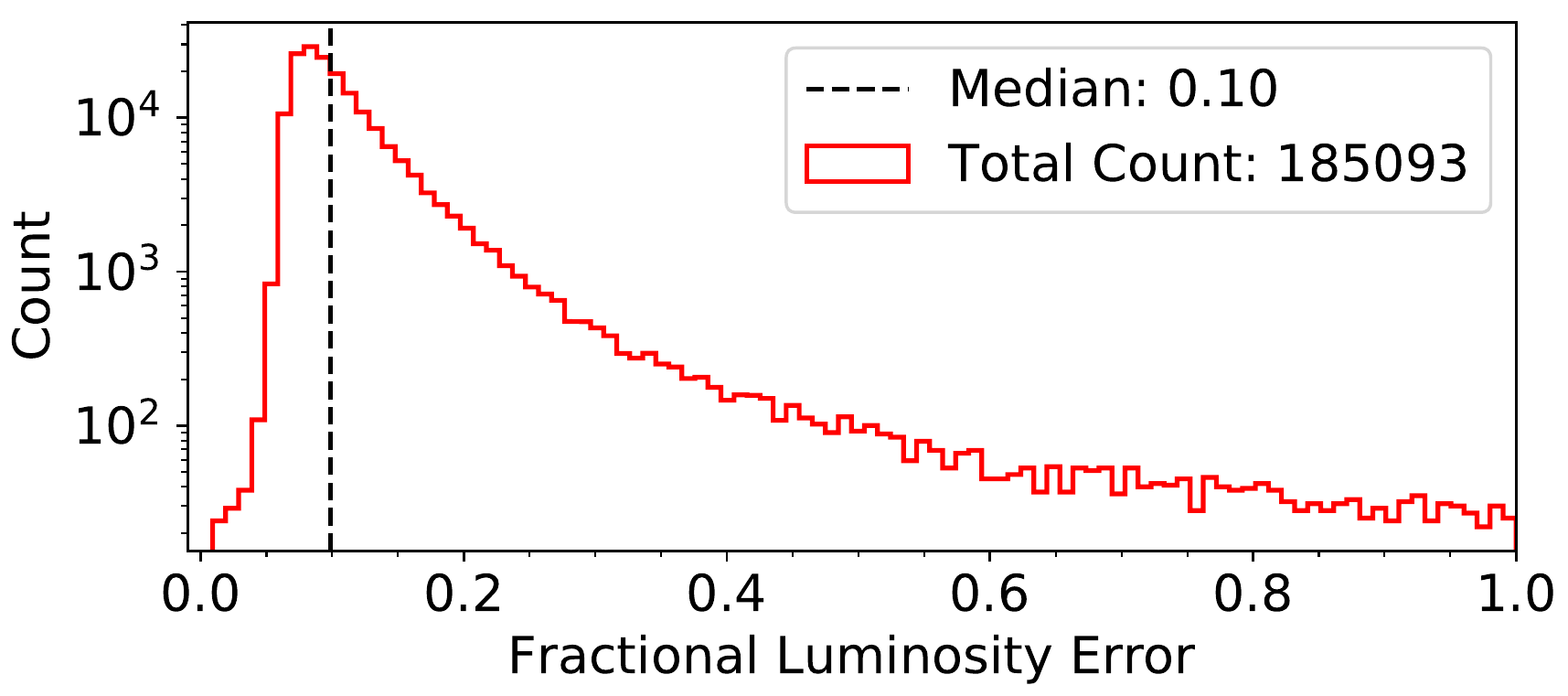}
\includegraphics[width=0.50\textwidth]{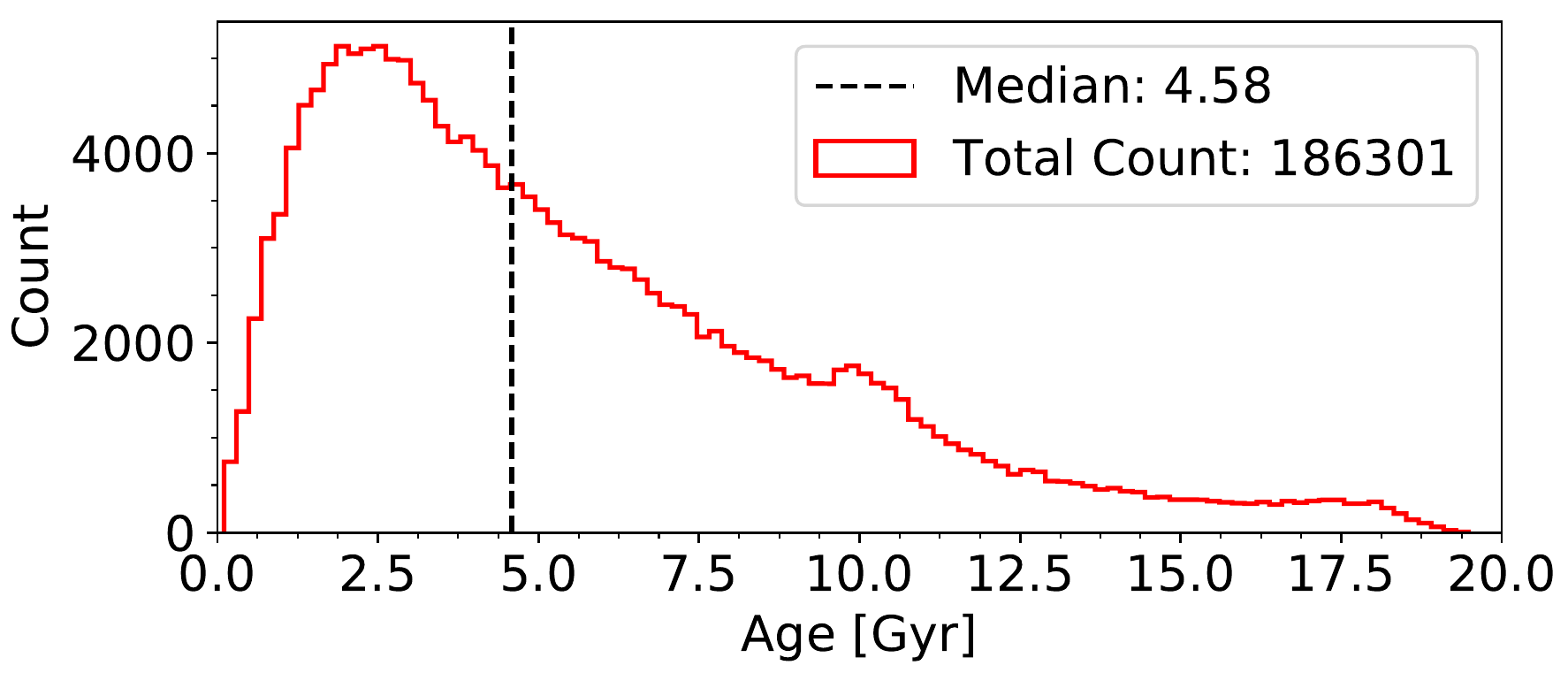}
\includegraphics[width=0.50\textwidth]{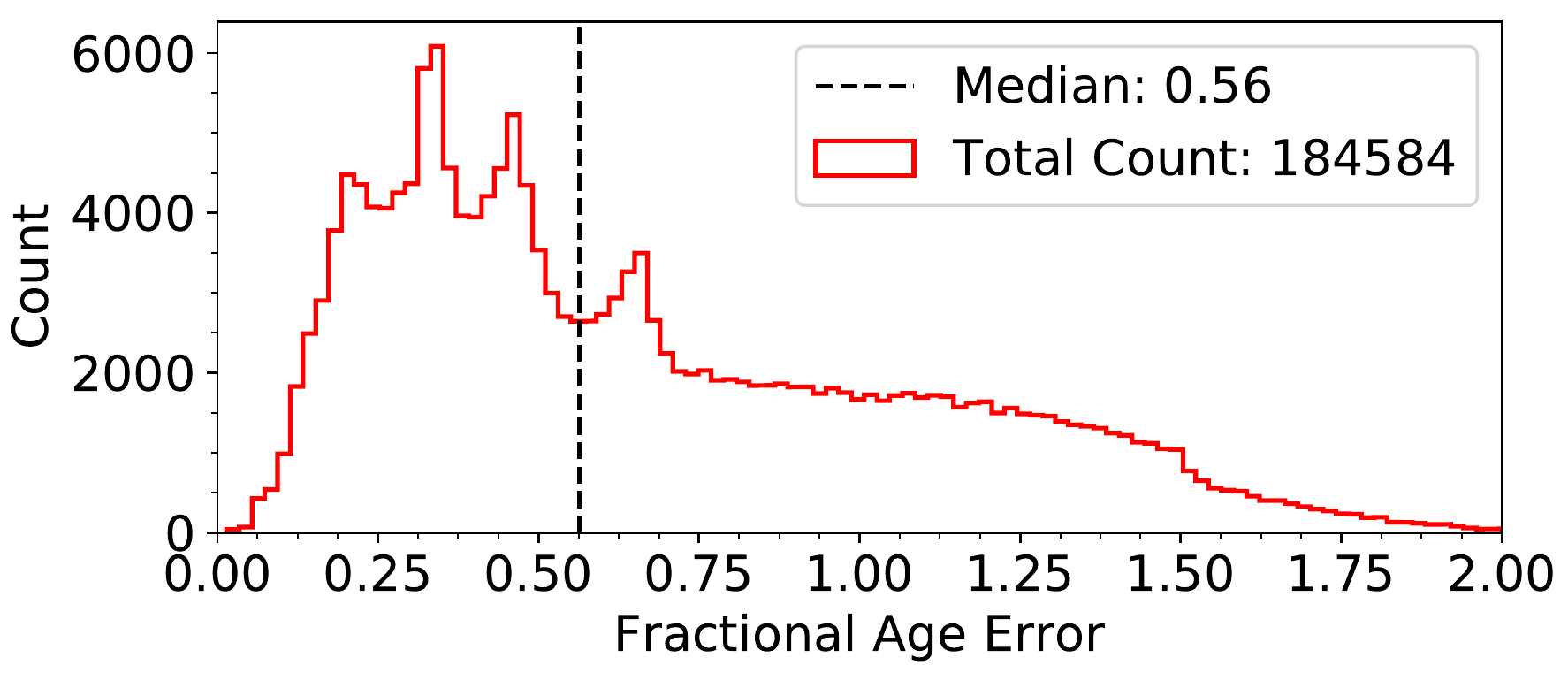}
\caption{Stellar parameter and uncertainty distributions from our catalog. The black dashed vertical lines illustrate the median value for each parameter, the value of which is given in the legend belonging to each plot. In addition, the total number of stars in displayed in each histogram is provided in the legend. These numbers vary due to choices in parameter cutoffs, and they are usually smaller than the total number of stars presented here, \nstars. Some histograms have logarithmic scaling on the x and/or y-axes.} 
\label{fig:StellarHist}
\end{figure*}

Figure \ref{fig:StellarHist} contains the remaining important parameters we derived for all \kep\ stars. Because many of our output posteriors are asymmetric, we compute our fractional errors by taking the maximum of the upper and lower uncertainties and then divide that maximum by the computed value for that parameter.

The first row of Figure \ref{fig:StellarHist} contains stellar masses. We see that the mass distribution peaks at 1\,\mdot, with a median that is slightly super-solar at 1.06\,\mdot. The fractional uncertainty distribution in mass peaks at $\approx$\,7\%, which is half that reported in \cite{Mathur2017}. The peak of fractional mass errors close to 2\% corresponds to the \cite{Mann2019} empirical $M_K$--mass uncertainties by design (\S\ref{sec:modelgrid}).

The second, third, and fourth rows of Figure \ref{fig:StellarHist} display the distributions of stellar radii, mean stellar densities ($\rho_\star$), and luminosities, respectively and their uncertainties. Each histogram peaks near solar values and plateaus towards larger radii and luminosities and smaller densities due to subgiant contamination. The more narrow peaks that occur around 11\,\rdot, 10$^{-3}$\,$\rho_\odot$, and 60\,\ldot\ represent the red clump. The uncertainty distributions for stellar radius, density, and luminosity peak at $\approx$\,3\%, $\approx$\,10\%, and $\approx$\,8\% and have a median of $\approx$\,4\%, 13\%, and 10\%, respectively. Each has a broad tail to larger fractional uncertainties, which is dependent mostly on the precision of the parallax from \gaia\ DR2. \teff\ errors are held fixed to $\approx$\,2\% as described in \S\ref{sec:modelgrid}. The 13\% median fractional error in $\rho_\star$ represents a factor of $\approx$\,4 improvment over the previous \kep\ Stellar Properties Catalog's fractional uncertainties \citep{Mathur2017}. These precise $\rho_\star$ values will be a critical input for refitting \kep\ transits.

The fifth row of Figure \ref{fig:StellarHist} contains stellar ages for the entire sample of \kep\ stars. The median value of 4.58\,Gyr is close to solar. The distribution peaks around 2.5\,Gyr and gradually falls off to larger ages. There is a bump at 10\,Gyr, half the age of the grid, where most of the M-dwarfs fall. This occurs because H-R diagram constraints are essentially uninformative for M-dwarfs given that even the most massive M-dwarfs have main sequence lifetimes $>$50\,Gyr, over twice the maximum age of our grid. Encouragingly, the distribution also qualitatively matches the red giant asteroseismology-derived age distributions in \cite{Aguirre2018} and \cite{Pinsonneault2018}, as well as the rotation-based ages in \cite{Claytor2019} and the Galactic Archaeology with HERMES--\gaia\ ages in \cite{Buder2019}.

The right histogram displaying the fractional age errors has a median fractional age uncertainty of 56\%. The peaks in the histogram represent various areas of parameter space where maximum fractional age errors are common. The first peak occurring at slightly less than 0.25 fractional age errors is one that corresponds to $\approx$\,0.9--1.3\,\mdot\ subgiants and $\approx$\,1.4--2.0\,\mdot\ TAMS \kep\ stars. The second, largest peak at $\approx$\,0.35 corresponds to the age uncertainties of (1) highest mass stars ($\gtrsim$\,1.3\,\mdot) on the main sequence, (2) intermediate mass stars ($\approx$\,1.3--1.7\,\mdot) at the TAMS, and (3) low mass stars ($\approx$\,0.7--1.3\,\mdot) on the subgiant branch, TAMS, and on the upper edge of the grid. Highlighting these stars in the H-R diagram outlines the main sequence turn-off ``hook''. The third peak, which occurs just below 0.5 includes (1) high mass stars ($\gtrsim$\,1.2\,\mdot) on the main sequence and (2) low mass stars ($\approx$\,0.7--1.4\,\mdot) at the TAMS and at the maximum ages within our grid. Finally, the fourth peak occurs at 0.65 fractional age errors because of the M-dwarfs and their uninformative ages. M-dwarfs do not evolve at all in 20\,Gyr, and hence have flat age posteriors with medians at $\approx$\,10\,Gyr and 1$\sigma$ uncertainties between 6 and 7\,Gyr. From there, the distribution smoothly decreases until 1.5 fractional age errors. Solar type stars at the ZAMS do not produce fractional age errors larger than 1.5 due to grid edge effects and the typical observational uncertainties, resulting in the sudden dip in the age distribution. Following this dip, the number of stars with larger and larger fractional age errors declines gradually.

\subsection{\teff\ Comparison to the DR25 \kep\ Stellar Properties Catalog} \label{sec:teffcomp}

\begin{figure}
\resizebox{\hsize}{!}{\includegraphics{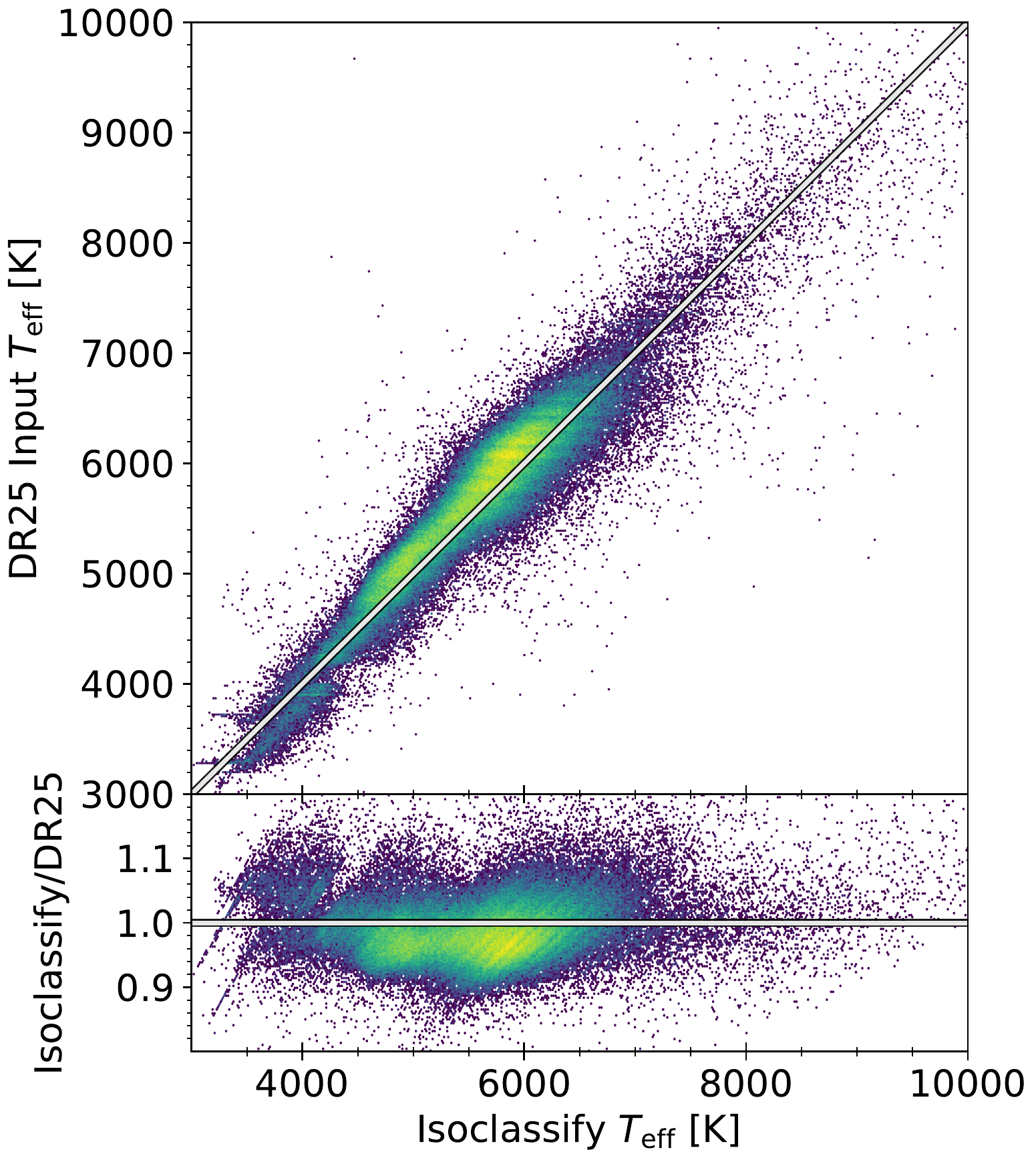}}
\caption{Comparison of \teff\ of the inputs to the DR25 \kep\ Stellar Properties Catalog \citep{Mathur2017} and the \teff\ derived in this paper. The colors represent the logarithmic density of points. The white and black line is the 1:1 comparison between DR25 \teff\ and our derived \teff. The bottom panel shows the ratio between DR25 stellar \teff\ and our stellar \teff.} \label{fig:teffcomp}
\end{figure}

Figure \ref{fig:teffcomp} shows a comparison of stellar \teff\ in the DR25 stellar properties catalog \citep{Mathur2017} to those derived in this paper. The distribution approximately tracks the 1:1 line. Our \teff\ are offset by --3\% for giant stars, +5\% for M-dwarfs, $<$\,1\% for late-K-dwarfs, --3\% for early K-dwarfs, --3\% for solar type stars, +1\% for F-dwarfs, and +10\% for A stars. For stars with \teff\,$>$\,10,000\,K, our \teff\ are $>$\,10\% larger. The majority of \cite{Mathur2017} \teff\ come from \cite{Pinsonneault2012}, which used KIC extinction values based on a simple extinction model \citep{brown11}. These extinction values were later shown to be overestimated \citep{Rodrigues2014}. In addition, the extinction used by \cite{Pinsonneault2012} does not account for each star's distance, where stars that are farther away will experience more extinction.

Due to reddening's increasing effect over longer distances, the \cite{Green2019} reddening map can account for the major differences that we see from the most distant, hottest stars down to the closer F-dwarfs. The solar-type stars and early K-dwarfs experience slightly less extinction than predicted \citep{Rodrigues2014}. Similarly, \cite{Huber2017} found that after accounting for the underestimated metallicity and overestimated extinction values used by \cite{Pinsonneault2012}, the \teff\ scales should be cooler by --20 to --65\,K. This cooler \teff\ scale brings us closer to the spectroscopic \teff\ \citep[see Figure 13 in][]{Pinsonneault2012}. However, the M-dwarfs are still too hot given their 2\% fractional errors while the giants are too cool. Much of this is likely due to the systematic issues displayed in Figure \ref{fig:teffint}, where the M-dwarfs are too hot and the giants exhibit a strong trend, likely created by systematic errors in color transformations.

The \teff\ gap at $\approx$\,4200\,K is visible in the \cite{Mathur2017} data. In addition, we observe banding structures which are visible as horizontal lines in the top plot and diagonal lines in the bottom residuals plot. This structure is an artifact in the input effective temperatures in \cite{Mathur2017}. It is unclear where exactly this banding comes from, but it appears to be dependent on a set of models, as the peaks are evenly spaced every $\approx$\,100\,K.

\section{Guidelines for Catalog Use} \label{sec:guidelines}

Our catalog includes multi-parameter solutions to \nstars\ \kep\ stars. In this section, we provide important guidelines, caveats, and limitations for the reader to implement/consider when utilizing this catalog:

\begin{itemize}
	\item Our input catalog contains 186,548 stars, while our output catalog contains \nstars\ stars. This is because the input parameters of 247 stars were too far removed from our grid of MIST models. Another 1543 stars have goodness-of-fit (GOF) parameters less than our threshold (GOF\,$<$\,0.99 in Table \ref{tab:output}). These stars should be used with caution.
	
	\item We only use a single model grid. Thus, differences due to input physics in model grids are not captured in the reported uncertainties.
	
	\item Our output metallicities ([Fe/H] in Table \ref{tab:output}) for the 120,000 stars constrained by the \kep\ field's 0.2\,dex dispersion solar metallicity prior should be treated with caution. The remaining 66,000 stars have spectroscopic metallicity constraints with 0.15\,dex uncertainty. Both sets of stars have large metallicity uncertainties.
	
	\item We do not treat (likely) binaries differently in our isochrone fitting analysis. We amend the 2MASS $K_s$-band magnitudes where possible, but do not modify the input observables or output stellar parameters of stars with large RUWE (RUWE\,$>$\,1.2 in Table \ref{tab:input}). These large RUWE stars and other likely binaries should be removed or treated with caution.
	
	\item We systematically overestimate M-dwarf \teff\ by $\approx$\,2\%, while our giant \teff\ exhibit a strong trend compared to interferometric determinations. This will affect our estimates of masses and radii for both M-dwarfs and giants, and giant masses, in particular, are extremely sensitive to metallicity and \teff, both of which are not constrained well in our catalog. Gaidos et al. (in prep) and APOKASC catalogs \citep{Serenelli2017,Pinsonneault2018} will provide better and more reliable parameters for \kep\ M-dwarfs and giants, respectively.
	
	\item We flag stellar ages which we deem unreliable (GOF\,$<$\,0.99 in Table \ref{tab:output}) or uninformative (TAMS\,$>$\,20\,Gyr in Table \ref{tab:output}) with asterisks, resulting in 14\% of catalog stars with suspect ages. We still provide the median and 1$\sigma$ confidence intervals for posterity, but these are stars whose ages cannot be constrained by our analysis. Due to degeneracies between stellar age and stellar metallicity, our most reliable stellar ages are dwarfs with spectroscopic metallicities ([Fe/H] constrained in Table \ref{tab:input}).
	
	\item We also caution against the use of our giant ages, given their strong dependence on stellar mass, which is strongly dependent on \teff\ and metallicity. For more reliable ages for many of the giants included in this catalog, see \cite{Serenelli2017,Pinsonneault2018}.
	
\end{itemize}

\section{Stellar Parameter Comparisons for Noteworthy \kep\ Systems} \label{sec:KOIcomp}

\begin{figure}
\resizebox{\hsize}{!}{\includegraphics{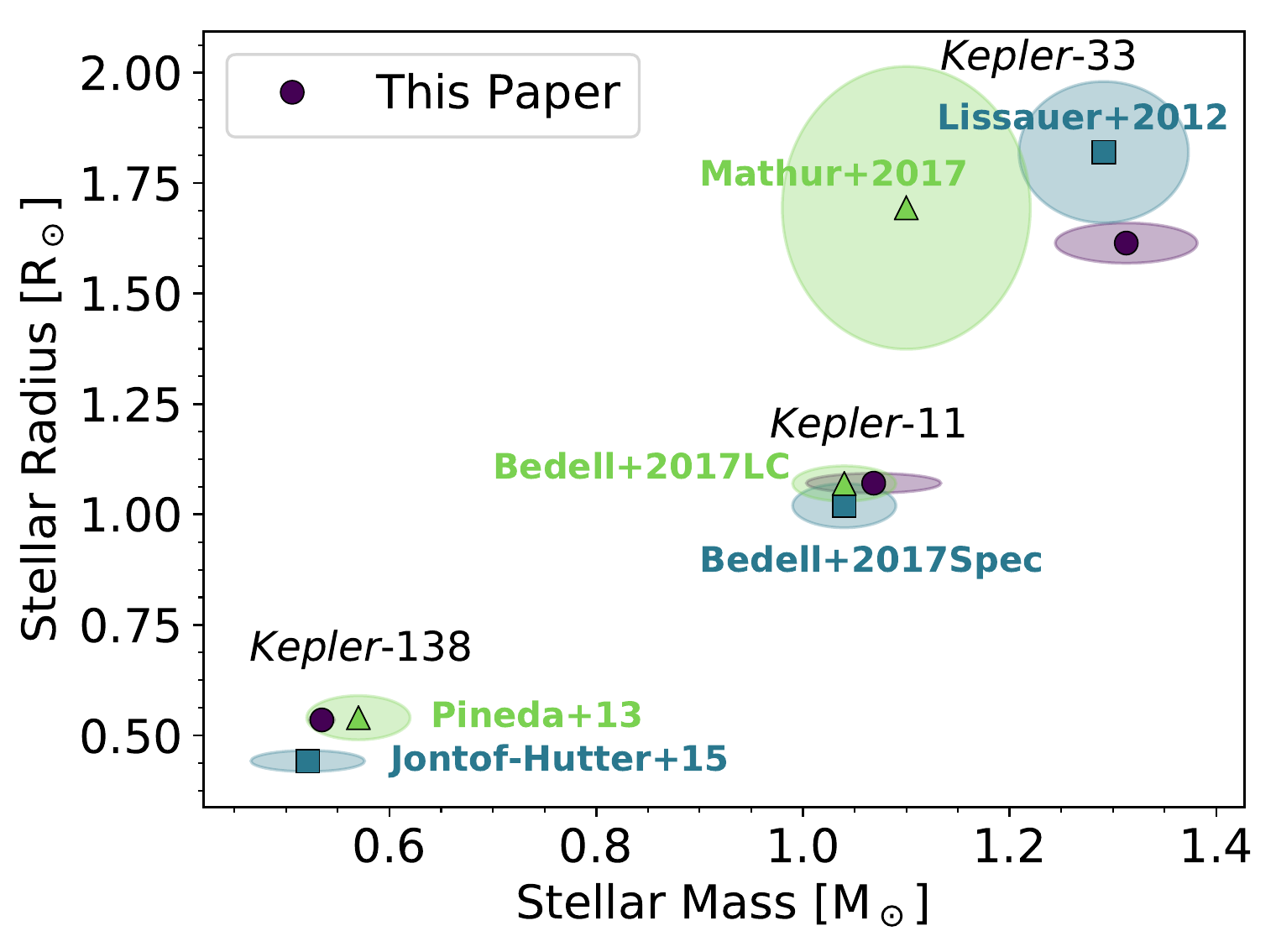}}
\caption{Stellar mass and radius comparisons for three particular \kep\ systems with stellar parameters at tension in the past. Points are colored and marked according to the source of their information, and the individual systems are labeled accordingly. Plum points and the purple 1$\sigma$ error ellipses are values and uncertainties determined from the analysis discussed above, while teal squares and green triangles and their respective error ellipses are taken from the literature. \kep-11 was investigated in \cite{Bedell2017}, \kep-33 in \cite{lissauer12} and \cite{Mathur2017}, and \kep-138 in \cite{Pineda2013} and \cite{Jontof2015}.}
\label{fig:stellarparcomp}
\end{figure}

To demonstrate the effectiveness of our catalog, we take a look at a few \kep\ systems that had stellar radius and mass estimates that were in tension before \gaia\ DR2. Figure \ref{fig:stellarparcomp} plots the stellar radius and mass measurements from a variety of sources for these systems. A more thorough investigation of updated planet radii of the \kep\ sample will be presented in a companion paper (Berger et al., in prep).

\subsection{\kep-11} \label{sec:kep11}

\kep-11 hosts six low density planets and was one of the first multiplanet systems discovered by \kep\ \citep{lissauer11}. The host star was analyzed most recently in \cite{Bedell2017}, where it was classified as a solar twin. \cite{lissauer13} also investigated the host's stellar properties by using transit timing variations (TTVs) to determine the star's density. \cite{Bedell2017} performed two analyses on \kep-11:  1) a spectroscopic determination of stellar parameters, and 2) a photodynamical light curve analysis. The spectroscopic analysis led to an estimation of $R_\star$\,$\approx$\,1.02\,$\pm$\,0.03\,\rdot\ and $M_\star$\,$\approx$\,1.04\,$\pm$\,0.01\,\mdot based on both Yonsei-Yale and Dartmouth isochrones. Keeping the stellar mass fixed at 1.04\,\mdot, the photodynamical analysis yielded a stellar radius of $1.07^{+0.04}_{-0.01}$\,\rdot. \cite{Bedell2017} then computed the mean stellar density for each of the methods, finding that they were at tension. The photodynamical analysis produced a mean stellar density in agreement with \cite{lissauer13}, while the spectroscopic analysis did not agree.

Our results (purple, center) appear to be in better agreement with the photodynamical analysis of the \kep-11 lightcurve, and hence also with the prediction of the stellar density computed in \cite{lissauer13}. While our reported stellar mass is greater than the mass derived in both the lightcurve and spectroscopic analysis in \cite{Bedell2017}, the 1$\sigma$ uncertainty includes the \cite{Bedell2017} 1.04\,\mdot\ estimates. \gaia\ DR2 parallaxes provide the strongest constraints on the stellar radius, which is in good agreement with the photodynamical and TTV analyses of the \kep-11 lightcurve, providing further evidence for the larger radius predicted by these methods.

\subsection{\kep-33} \label{sec:kep33}

\kep-33, investigated in \cite{lissauer12}, has five planetary companions, all between the sizes of 1.5--6.0\,$\mathrm{R_\oplus}$. In \cite{lissauer12}, both the mass and age posteriors are bimodal due to the star's location on the H-R diagram, close to the main-sequence turn-off and subgiant branch. \cite{lissauer12} reports a mass of 1.29\,$\pm$\,0.08\,\mdot\ and a radius of 1.82\,$\pm$\,0.16\,\rdot based on the Yonsei-Yale \citep{yi01,Kim2002,Yi2003,Demarque2004} isochrone placement of the spectroscopically derived parameters (teal square). However, while \cite{Mathur2017} used the spectroscopic parameters of \cite{lissauer12}, \cite{Mathur2017} used Dartmouth Stellar Evolution Database \citep[DSEP,][]{dotter08} models to derive $R_\star$\,=\,1.69\,$\pm$\,0.32\,\rdot\ and $M_\star$\,=\,1.10\,$\pm$\,0.12\,\mdot (green triangle). Therefore, the stellar mass tension is result of different stellar model grids.

Our result is more than 1$\sigma$ discrepant from the \cite{Mathur2017} mass and the \cite{lissauer12} radius. Because the stellar masses derived by \cite{lissauer12} and \cite{Mathur2017} are model grid-dependent, we focus on the discrepancy in stellar radius. The stellar radius in \cite{lissauer12} is mostly constrained by the spectroscopic estimate of \logg. However, spectroscopic \logg\ values are often degenerate with \teff\ and [Fe/H] \citep{Torres2012}. Alternatively, parallaxes, which constrain distances and hence radii, provide a more straightforward approach to determining stellar radii. Hence, we are confident in our parallax-constrained properties of \kep-33.

\subsection{\kep-138} \label{sec:kep138}

Investigated in both \cite{Pineda2013} and \cite{Jontof2015}, \kep-138 has three planetary companions, one of which was determined to be a Mars-sized planet in \cite{Jontof2015}. The two results are at tension, largely because different methods were used to determine the parameters. \cite{Pineda2013} utilized an empirical relation that determined absolute $K_s$ magnitudes from the equivalent widths of molecular lines (TiO, VO) and photometric colors. These absolute magnitudes were then converted to masses using an absolute magnitude-mass relation from \cite{Delfosse2000} and then to masses using the mass-radius relation of \cite{boyajian12b}. Alternatively, \cite{Jontof2015} computed the stellar parameters of \kep-138 by fitting the light curve constraint of $\rho_\star$ as well as the spectroscopic constraints of \teff\ and [Fe/H] of \cite{muirhead12b} to Dartmouth Stellar Evolution models.

According to our analysis, \kep-138 has a mass of 0.53\,\mdot\ and a radius of 0.54\,\rdot. Because this particular star is an M-dwarf, we caution that our mass is suspect given the systematically overestimated \teff\ of M-dwarfs demonstrated in Figure \ref{fig:teffint} above. For instance, if we corrected \kep-138's \teff\ by the 70\,K systematic offset seen for interferometric stars, we would compute $\approx$\,0.5\,\mdot\ and $\approx$\,0.5\,\rdot\ for the star's mass and radius, which would agree with neither \cite{Pineda2013} nor \cite{Jontof2015}. We note that both the radius and the mass error bars are smaller than the marker size.

Our radius and mass estimates for \kep-138 agree with the reported masses and radii in \cite{Pineda2013}, while they disagree with those reported in \cite{Jontof2015}. Although \cite{Jontof2015} cites a few potential inaccuracies of the \cite{Pineda2013} results, it appears that our solution breaks the tension in favor of the \cite{Pineda2013} result. Possible reasons could be inaccuracies in the light curve modeling or photodynamical modeling for the determination of $\rho_\star$. Any inaccuracies in the mean stellar density will scale as $R_\star^{-3}$ while only linearly in mass, which explains the large discrepancy in radius as compared to the one in mass. However, due to the systematic offset in M-dwarf \teff\ seen in Figure \ref{fig:teffint}, we caution against drawing any strong conclusions for the properties of \kep-138.

\section{Summary and Conclusions} \label{sec:conc}

We presented a re-classification of stellar parameters -- \teff, masses, radii, luminosities, densities, surface gravities, ages, and metallicities -- for \nstars\ stars observed by the \kep\ mission by combining \gaia\ DR2 parallaxes and spectroscopic metallicities with calibrated KIC \citep{brown11} and KIS $g$ \citep{Greiss2012} and visual-binary-de-contaminated 2MASS $K_s$ \citep{skrutskie06} photometry. We utilized a custom-interpolated set of MIST models \citep{choi16} and \texttt{isoclassify} \citep{Huber2017} to derive stellar parameters. Our main results are as follows:

\begin{itemize}
   \item We determine parameters for \nstars\ stars. The median (fractional) precisions of our \teff, \logg, radii, masses, mean stellar densities, luminosities, and ages are 112\,K, 0.05\,dex, 4\%, 7\%, 13\%, 10\%, and 56\%, respectively.
   
   \item We provide the first \kep\ Stellar Properties Catalog with a homogeneous \teff\ scale. M-dwarf \teff\ may be 75\,K hotter than similar stars with interferometry, and our FGK-dwarf \teff\ are cooler than \cite{Mathur2017} by $\approx$\,110\,K.
   
   \item We derive a median \kep\ target age of $\approx$\,4.6\,Gyr. Our ages are in good agreement with cluster and asteroseismic ages, where we find that our median age is 3\% larger than the asteroseismic estimate with a scatter of 29\%. We caution that 14\% of our ages are uninformative, due to the constraints of isochrone fitting for low-mass stars. Our ages are most reliable for both the most massive stars and those on the subgiant branch.

    \item We provide the first \kep\ Stellar Properties Catalog which attempts to account for the binarity of all \kep\ stars when performing isochrone fits to absolute $K_s$-band photometry. In addition, we find that at least 12\% of the \kep\ sample is affected by binary companions. When holding age and metallicity constant, we observe that age biases of companions are functions of both primary and secondary mass; we find binary companions will bias the ages of lower-main sequence stars by as much as 10\,Gyr (where age uncertainties are $\gtrsim$\,6\,Gyr), and higher mass stars by a few Gyr (where age uncertainties are $\lesssim$\,1\,Gyr).
    
    \item We derive accurate and precise stellar masses and radii for three \kep\ systems with tension in their reported parameters based on previous analyses. Our results typically break the tension and favor one result over another, although we suggest the reader carefully consider the methods used in each analysis before drawing any strong conclusions.
\end{itemize}

All of the homogeneous parameters reported here will prove useful for future \kep\ exoplanet occurrence rate computations, as homogeneous treatment for all stars ensures that both the host star and field star parameters are considered similarly. For instance, \cite{Bryson2019} utilized the parameters presented in this work to investigate the DR25 catalog's reliability and completeness.

In addition, the masses and ages presented here provide important constraints for \kep\ exoplanet host stars. The masses determined here will allow us to constrain the stellar mass dependence of the planet radius gap \citep{Fulton2018,Gupta2019,Wu2019}. Stellar ages also have interesting implications for exoplanets. Previous analyses have hinted at age-dependent effects on the radii of small exoplanets, particularly those at or near the gap, for subsamples of the \kep\ exoplanets \citep{Fulton2017,Mann2017,Berger2018}. With our stellar mass and age constraints for the entire \kep\ exoplanet host sample, we will investigate stellar mass and age-dependent exoplanet trends in our companion paper (Berger et al. 2019b, in prep). Ultimately, we look forward to future investigations which will discover both new features about the \kep\ sample and confirm previous results, continuing the legacy of the \kep\ telescope well beyond its final observation.

\begin{deluxetable*}{ccccccccccccc}
\tabletypesize{\scriptsize}
\tablenum{1}
\tablewidth{0pt}
\tablecolumns{12}
\tablecaption{\gaia-\kep\ Stellar Input Parameters}
\tablehead{
\colhead{KIC ID} & \colhead{$g$ [mag]} & \colhead{$\sigma_{g}$ [mag]} & \colhead{$K_s$ [mag]} & \colhead{$\sigma_{K}$ [mag]} & \colhead{$\pi$ [mas]} & \colhead{$\sigma_{\pi}$ [mas]} & \colhead{[Fe/H]} & \colhead{$\sigma_{\mathrm{[Fe/H]}}$} & \colhead{RUWE} & \colhead{\# Companions} & \colhead{$K_s$ Prov} & \colhead{Ev.\ State}}
\def\arraystretch{1.0}
\startdata
757076&12.351&0.020&9.559&0.017&1.524&0.048&&&0.947&&&\\
757099&13.704&0.020&11.094&0.018&2.708&0.027&&&2.173&&&\\
757137&10.052&0.028&6.722&0.017&1.753&0.025&&&0.913&&&RGB\\
757280&12.133&0.020&10.627&0.018&1.214&0.022&&&0.870&&&\\
757450&15.895&0.020&13.118&0.029&1.199&0.026&0.229&0.15&1.030&&&\\
891901&13.631&0.020&11.928&0.020&0.857&0.116&&&7.356&&&\\
891916&15.354&0.020&13.076&0.026&0.453&0.215&&&8.712&&&\\
892010&12.617&0.021&9.041&0.017&0.541&0.024&&&1.014&&&\\
892107&13.131&0.023&10.163&0.017&1.064&0.023&&&0.940&&&clump\\
892195&14.340&0.020&11.814&0.019&2.080&0.017&&&1.121&&&\\
892203&14.033&0.020&11.950&0.020&1.802&0.016&&&1.245&&&\\
892376&15.521&0.020&10.721&0.015&3.038&0.360&&&16.182&&&\\
892667&13.424&0.020&11.818&0.020&0.851&0.015&&&0.867&&&\\
892675&13.860&0.020&11.940&0.020&1.711&0.014&&&1.065&&&\\
892678&12.536&0.021&10.580&0.018&1.024&0.026&&&0.985&1.000&BinaryCorr&\\
892713&12.466&0.021&10.509&0.017&0.960&0.023&&&0.987&1.000&BinaryCorr&\\
\enddata
\tablecomments{KIC ID, $g$-mag, $K_s$-mag, parallax, metallicity, RUWE flag, number of companions within 4'' detected by \gaia, $K_s$-mag flag indicating potential corrections compared to 2MASS $K_s$ (empty rows indicate no correction), and giant branch evolutionary state flag from \cite{vrard16} and \cite{Hon2018} as input parameters for our sample of \nstarsinput\ \kep\ stars. A subset of our input parameters is provided here to illustrate the form and format. The full table, in machine-readable format, can be found online.} \label{tab:input}
\end{deluxetable*}

\begin{deluxetable*}{cllllllcrrrrr}
\tabletypesize{\scriptsize}
\tablenum{2}
\tablewidth{0pt}
\tablecolumns{13}
\tablecaption{\gaia-\kep\ Stellar Output Parameters}
\tablehead{\colhead{KIC ID} & \colhead{\teff\ [K]} & \colhead{\logg\ [dex]} & \colhead{[Fe/H]} & \colhead{$M_\star$ [$\mathrm{M_\odot}$]} & \colhead{$R_\star$ [$\mathrm{R_\odot}$]} & \colhead{$\log{\rho_\star}$ [$\mathrm{\rho_\odot}$]} & \colhead{$\log{L_\star}$ [$\mathrm{L_\odot}$]} & \colhead{Age [Gyr]} & \colhead{Distance [pc]} & \colhead{$A_V$ [mag]} & \colhead{GOF} & \colhead{TAMS [Gyr]}}
\def\arraystretch{1.0}
\startdata
757076&5052$^{+103}_{-86}$&3.37$^{+0.07}_{-0.08}$&-0.14$^{+0.16}_{-0.19}$&1.40$^{+0.18}_{-0.22}$&4.00$^{+0.14}_{-0.15}$&-1.67$^{+-2.38}_{--2.40}$&0.98$^{+-0.04}_{--0.10}$&2.5$^{+1.8}_{-0.7}$&651$^{+22}_{-21}$&0.37&1.0000&2.7\\
757099&5364$^{+102}_{-84}$&4.32$^{+0.04}_{-0.03}$&0.08$^{+0.14}_{-0.13}$&0.87$^{+0.05}_{-0.04}$&1.07$^{+0.02}_{-0.02}$&-0.15$^{+-1.13}_{--1.25}$&-0.07$^{+-1.23}_{--1.29}$&15.2$^{+3.0}_{-4.0}$&367$^{+7}_{-6}$&0.34&1.0000&17.2\\
757137&4628$^{+84}_{-76}$&2.39$^{+0.08}_{-0.09}$&-0.11$^{+0.15}_{-0.17}$&1.67$^{+0.31}_{-0.30}$&13.59$^{+0.32}_{-0.33}$&-3.19$^{+-3.83}_{--3.93}$&1.88$^{+0.74}_{-0.64}$&1.5$^{+1.1}_{-0.6}$&568$^{+12}_{-11}$&0.34&1.0000&1.7\\
757280&6856$^{+144}_{-139}$&3.83$^{+0.03}_{-0.03}$&-0.03$^{+0.21}_{-0.11}$&1.71$^{+0.09}_{-0.09}$&2.61$^{+0.07}_{-0.07}$&-1.02$^{+-2.01}_{--2.06}$&1.13$^{+0.05}_{-0.00}$&1.2$^{+0.2}_{-0.2}$&822$^{+19}_{-21}$&0.50&1.0000&1.6\\
757450&5301$^{+111}_{-103}$&4.43$^{+0.05}_{-0.04}$&0.24$^{+0.13}_{-0.13}$&0.91$^{+0.06}_{-0.06}$&0.96$^{+0.03}_{-0.03}$&0.01$^{+-0.85}_{--0.91}$&-0.18$^{+-1.24}_{--1.28}$&9.5$^{+5.4}_{-5.1}$&829$^{+24}_{-23}$&0.46&1.0000&16.1\\
891901&6350$^{+130}_{-131}$&3.96$^{+0.10}_{-0.09}$&0.02$^{+0.15}_{-0.14}$&1.41$^{+0.12}_{-0.12}$&2.02$^{+0.28}_{-0.26}$&-0.80$^{+-1.16}_{--1.35}$&0.78$^{+0.30}_{-0.17}$&2.2$^{+0.7}_{-0.5}$&1122$^{+156}_{-146}$&0.34&1.0000&2.9\\
891916&5650$^{+131}_{-137}$&4.13$^{+0.22}_{-0.25}$&0.02$^{+0.15}_{-0.17}$&1.01$^{+0.16}_{-0.11}$&1.35$^{+0.59}_{-0.36}$&-0.46$^{+-0.37}_{--0.68}$&0.24$^{+0.25}_{--0.08}$&7.6$^{+3.7}_{-3.3}$&1193$^{+515}_{-322}$&0.36&1.0000&9.7\\
892010&4555$^{+141}_{-92}$&2.30$^{+0.16}_{-0.12}$&-0.02$^{+0.16}_{-0.20}$&1.71$^{+0.71}_{-0.40}$&15.19$^{+0.77}_{-0.76}$&-3.32$^{+-3.65}_{--3.92}$&1.96$^{+1.07}_{-0.99}$&1.4$^{+1.7}_{-0.8}$&1832$^{+87}_{-87}$&0.37&1.0000&1.6\\
892107&4894$^{+83}_{-85}$&3.24$^{+0.08}_{-0.10}$&-0.05$^{+0.14}_{-0.15}$&1.24$^{+0.21}_{-0.24}$&4.41$^{+0.13}_{-0.12}$&-1.84$^{+-2.53}_{--2.51}$&1.01$^{+-0.11}_{--0.15}$&4.1$^{+4.5}_{-1.7}$&936$^{+25}_{-24}$&0.28&1.0000&4.3\\
892195&5333$^{+101}_{-84}$&4.37$^{+0.04}_{-0.03}$&0.07$^{+0.14}_{-0.13}$&0.86$^{+0.06}_{-0.04}$&1.00$^{+0.02}_{-0.02}$&-0.07$^{+-1.02}_{--1.17}$&-0.14$^{+-1.31}_{--1.36}$&14.3$^{+3.5}_{-4.6}$&479$^{+8}_{-8}$&0.25&1.0000&18.0\\
892203&5712$^{+108}_{-105}$&4.39$^{+0.04}_{-0.04}$&0.01$^{+0.15}_{-0.15}$&0.97$^{+0.07}_{-0.07}$&1.04$^{+0.02}_{-0.02}$&-0.07$^{+-0.99}_{--1.02}$&0.02$^{+-1.14}_{--1.19}$&6.5$^{+4.1}_{-3.5}$&553$^{+10}_{-10}$&0.22&1.0000&11.3\\
892667&6704$^{+148}_{-128}$&3.95$^{+0.03}_{-0.04}$&0.01$^{+0.16}_{-0.17}$&1.55$^{+0.08}_{-0.09}$&2.17$^{+0.06}_{-0.06}$&-0.83$^{+-1.81}_{--1.84}$&0.94$^{+-0.15}_{--0.19}$&1.6$^{+0.3}_{-0.3}$&1171$^{+29}_{-29}$&0.50&1.0000&2.2\\
892675&5929$^{+108}_{-108}$&4.39$^{+0.04}_{-0.04}$&-0.02$^{+0.14}_{-0.16}$&1.04$^{+0.07}_{-0.08}$&1.08$^{+0.02}_{-0.02}$&-0.09$^{+-1.09}_{--1.06}$&0.12$^{+-1.05}_{--1.08}$&3.6$^{+3.2}_{-2.2}$&583$^{+11}_{-10}$&0.25&1.0000&8.5\\
892678&5890$^{+121}_{-114}$&3.57$^{+0.03}_{-0.03}$&0.02$^{+0.19}_{-0.22}$&1.58$^{+0.07}_{-0.06}$&3.38$^{+0.12}_{-0.10}$&-1.40$^{+-2.37}_{--2.41}$&1.10$^{+0.11}_{-0.01}$&1.8$^{+0.1}_{-0.1}$&967$^{+34}_{-17}$&0.25&1.0000&2.1\\
892713&6238$^{+123}_{-129}$&3.55$^{+0.07}_{-0.04}$&0.08$^{+0.21}_{-0.20}$&1.73$^{+0.22}_{-0.09}$&3.64$^{+0.12}_{-0.11}$&-1.45$^{+-2.20}_{--2.40}$&1.25$^{+0.25}_{--0.15}$&1.4$^{+0.1}_{-0.3}$&1033$^{+32}_{-29}$&0.50&1.0000&1.7\\
892718&5000$^{+97}_{-90}$&4.57$^{+0.03}_{-0.04}$&-0.08$^{+0.14}_{-0.13}$&0.78$^{+0.04}_{-0.05}$&0.76$^{+0.03}_{-0.03}$&0.25$^{+-0.77}_{--0.74}$&-0.49$^{+-1.46}_{--1.50}$&6.3*$^{+7.1*}_{-4.5*}$&874$^{+32}_{-31}$&0.31&1.0000&23.1\\
\enddata
\tablecomments{KIC ID, effective temperature, surface gravity, surface metallicity, stellar mass, stellar radius, density, luminosity, age, distance, $V$-magnitude extinction, combined likelihood goodness-of-fit (GOF), and terminal age of the main sequence (TAMS) parameters and their errors for \nstars\ \kep\ stars, output from our isochrone placement routine detailed in \S\ref{sec:methods}. Ages with asterisks are either those with uninformative posteriors (TAMS\,$>$\,20\,Gyr) or unreliable ages (GOF\,$<$\,0.99). Stars within Table \ref{tab:input} and not in this table have fewer than ten models within 4$\sigma$ of the input observables. A subset of our output parameters is provided here to illustrate the form and format. The full table, in machine-readable format, can be found online.} \label{tab:output}
\end{deluxetable*}

\acknowledgments
We gratefully acknowledge everyone involved in the \gaia\ and \kep\ missions for their tireless efforts which have made this paper possible. We thank Marc Pinsonneault, Jack Lissauer, Tim White, Dennis Stello, Sam Grunblatt, Lauren Weiss, Ashley Chontos, Erica Bufanda, Maryum Sayeed, Connor Auge, Vanshree Bhalotia, Nicholas Saunders, Michael Liu, Benjamin Boe, and Ehsan Kourkchi for helpful discussions in addition to feedback on the figures. In addition, we thank Jason Drury and Diego Godoy-Rivera for providing cluster membership for NGC 6791 and 6811, respectively. T.A.B. and D.H. acknowledge support by a NASA FINESST award (80NSSC19K1424) and the National Science Foundation (AST-1717000). D.H. acknowledges support from the Alfred P. Sloan Foundation. E.G. acknowledges support from NSF award AST-187215. E.G. was also supported as a visiting professor to the University of Göttingen by the German Science Foundation through DFG Research 644 Unit FOR2544 ``Blue Planets around Red Stars". J.T. acknowledges that support for this work was provided by NASA through the NASA Hubble Fellowship Grant \#51424 awarded by the Space Telescope Science Institute, which is operated by the Association of Universities for Research in Astronomy, Inc., for NASA, under contract NAS5-26555. This research was partially conducted during the Exostar19 program at the Kavli Institute for Theoretical Physics at UC Santa Barbara, which was supported in part by the National Science Foundation under Grant No. NSF PHY-1748958. This work has made use of data from the European Space Agency (ESA) mission {\it Gaia} (\url{https://www.cosmos.esa.int/gaia}), processed by the {\it Gaia} Data Processing and Analysis Consortium (DPAC, \url{https://www.cosmos.esa.int/web/gaia/dpac/consortium}). Funding for the DPAC has been provided by national institutions, in particular the institutions participating in the {\it Gaia} Multilateral Agreement. Guoshoujing Telescope (the Large Sky Area Multi-Object Fiber Spectroscopic Telescope LAMOST) is a National Major Scientific Project built by the Chinese Academy of Sciences. Funding for the project has been provided by the National Development and Reform Commission. LAMOST is operated and managed by the National Astronomical Observatories, Chinese Academy of Sciences. This publication makes use of data products from the Two Micron All Sky Survey, which is a joint project of the University of Massachusetts and the Infrared Processing and Analysis Center/California Institute of Technology, funded by the National Aeronautics and Space Administration and the National Science Foundation. This research has made use of NASA's Astrophysics Data System. This research was made possible through the use of the AAVSO Photometric All-Sky Survey (APASS), funded by the Robert Martin Ayers Sciences Fund. This research made use of the cross-match service provided by CDS, Strasbourg. This research has made use of the NASA Exoplanet Archive, which is operated by the California Institute of Technology, under contract with the National Aeronautics and Space Administration under the Exoplanet Exploration Program.

\vspace{5mm}

\software{\texttt{astropy} \citep{astropy},
		  \texttt{dustmaps} \citep{Green2018},
		  \texttt{GNU Parallel} \citep{Tange2018},
		  \texttt{isoclassify} \citep{Huber2017}, 
		  \texttt{Matplotlib} \citep{Matplotlib},
          \texttt{mwdust} \citep{bovy16}, 
          \texttt{Pandas} \citep{Pandas}, 
          \texttt{SciPy} \citep{Scipy}}

\bibliography{references}

\end{document}